\documentclass[a4paper,fleqn,titlepage,twocolumn,superscriptaddress,showkeys]{revtex4-2}
\usepackage[utf8]{inputenc}
\usepackage{graphicx}
\usepackage{amsmath}
\usepackage{amssymb}
\usepackage{times}
\usepackage[utf8]{inputenc}
\usepackage{nicefrac}
\usepackage{color}
\usepackage{xcolor}
\definecolor{darkblue}{rgb}{0.0, 0.18, 0.65}
\usepackage{hyperref}
\hypersetup{
     colorlinks=true,
     linkcolor=blue,
     filecolor=blue,
     citecolor = blue,      
     urlcolor=blue}
     
\usepackage{color}
\usepackage{subcaption}
\usepackage{dsfont}
\usepackage{xr} 
\usepackage[normalem]{ulem} 
\graphicspath{{figs/}}

\newcommand{\bal}{\tau}

\begin{document}


\title{Modelling echo chamber effects in signed networks}

\author{Antoine Vendeville}
\email{Corresponding author: antoine.vendeville@sciencespo.fr}
\affiliation{m\'edialab, Sciences Po, 75007 Paris, France}
\affiliation{Complex Systems Institute of Paris Île-de-France (ISC-PIF) CNRS, 75013 Paris, France}
\affiliation{Learning Planet Institute, Research Unit Learning Transitions, 75004 Paris, France}
\author{Fernando Diaz-Diaz}
\email{fernandodiaz@ifisc.uib-csic.es}
\affiliation{IFISC (UIB-CSIC), Institute for Cross-Disciplinary Physics and Complex Systems, Campus Universitat de les Illes Balears, 07122 Palma de Mallorca, Spain}

\date{\today}

\begin{abstract}
Echo chamber effects in social networks are generally attributed to the prevalence of interactions among like-minded peers. However, recent evidence has emphasized the role of hostile interactions between opposite-minded groups. We investigate the role of polarization, identified with structural balance, in the formation of echo chambers in signed networks. To do so, we generalize the Independent Cascade Model and the Linear Threshold Model to describe information propagation in presence of negative edges. Antagonistic connections do not disrupt the flow of information, but instead, alter the way information is framed. Our results show that echo chambers spontaneously emerge in balanced networks, but also in antibalanced ones for specific parameters. This highlights that structural polarization and echo chambers do not necessarily display a one-to-one correspondence, showing instead a complex and often counterintuitive interplay. The robustness of our results is confirmed with a complex contagion model and through simulations in different network topologies, including real-world datasets. 
\end{abstract}

\keywords{Echo chambers, Signed networks, Structural balance, Social networks}
\maketitle

\section{Introduction}
The study of echo chambers has been central in research about online social platforms. By continuously reinforcing pre-existing views \cite{baumann2020, baqir2023beyond}, these tightly knit communities of like-minded users exacerbate the polarization of opinions and in turn, the fragmentation of societies \cite{conover2011,halberstam2016,garimella2018,cinelli2021}. They have been observed at multiple levels, from the connections established by the users themselves, to personalised timelines of content curated by engagement-seeking algorithms \cite{bakshy2015,bailon2023}. 

In recent years however, this evidence has been increasingly challenged \cite{flaxman2016,dubois2018,yang2020,defranciscimorales2021,guess2021,tornberg2022}. Scholars have pointed out that there is communication between opposite-minded communities, often in the form of hostile interactions \cite{williams2015,tacchi2022,efstratiou2023}. Opposite-minded users may frame the same piece of information under different lights, reinforcing ideological divides while still allowing information to travel through partisan lines. Negative relations might be even more prevalent than positive ones for some users, such as journalists \cite{tacchi2022}. Similarly, recent research has highlighted the role of affective polarization, a phenomenon by which people experience positive feelings towards like-minded peers and negative feelings towards others \cite{iyengar2019origins,yarchi2020political}. Thus, it is primordial that we distinguish between friends and foes when studying online debates \cite{keuchenius2021}. 

The presence of hostile interactions among individuals is often modelled as a \textit{signed network} with positive and negative links \cite{zaslavsky1982signed}. 
The inclusion of negative links gives rise to novel phenomena, such as structural balance \cite{harary1953notion, cartwright1956structural}. A signed network is \textit{perfectly balanced} if it can be divided into two groups, so that all links between agents of the same group are positive and all links between agents of different groups are negative \cite{harary1953notion}. In other words, it contains two groups that are antagonistic towards one another \cite{neal2020, diaz2024signed}. While perfectly balanced networks are a mathematical idealization, studies have consistently shown a tendency towards structural balance in empirical networked systems of various types \cite{saiz2017evidence, kirkley2019balance, facchetti2011computing, gallo2024testing}. On the other hand, when all links within each group are mostly negative and links between groups are mostly positive, the network is said to be \textit{antibalanced} \cite{harary1957structural}. While not as common as balanced networks, antibalanced networks can emerge in historical, social and political contexts, like international relations during the Thirty Years' War \cite{wilson2012meaningless} or the U.S. Congress during the Carter and Reagan administrations (see \cite{neal2020} and Fig.~\ref{congress_stats}(a)). 

These findings have motivated the introduction of the notion of \textit{degree of balance}, with indices based on triangles \cite{gallo2024testing, neal2020}, random walks \cite{estrada2014walk, talaga2023polarization, diazdiaz2024mathematical} or frustration metrics \cite{aref2018measuring, fraxanet2024unpacking}, among others. Recent research has discovered intriguing connections between the degree of balance of social and political networks and their tendency towards affective \cite{fraxanet2024unpacking} and ideological \cite{neal2020} polarization. In essence, the degree of balance—an inherently structural property—is strongly tied to the social and political phenomenon of polarization, a connection we term \textit{structural polarization}. The degree of balance also impacts the evolution of various dynamical processes \cite{altafini2012, shi2019dynamics, diazdiaz2024mathematical}, including contagion models \cite{fan2012, yin2021, lee2023threshold}, random walks \cite{tian2024spreading}, diffusive dynamics \cite{altafini2013} and the voter model \cite{li2015}. Conversely, structural balance can spontaneously arise as a result of opinion dynamics on signed networks \cite{pham2020,pham2022}.

In this work, we add onto this literature by investigating the role of the degree of balance in the emergence of echo chambers. Doing so, we establish a correspondence between structural polarization and echo chamber effects, two distinct yet closely related phenomena. Our contribution is two-fold. First, we extend to signed networks two popular models of information propagation, the Independent Cascade Model and the Linear Threshold Model \cite{kempe2003, watts2002simple}. The key idea driving these extensions is that a single piece of information may be framed in different ways between, and within, antagonistic groups. Second, we use those models to analyze the key role played by the degree of balance in the emergence of echo chambers. Echoing empirical observations on the correlation between activity and polarisation \cite{vaccari2016,hills2019,baumann2020,hosseinmardi2021}, we find reinforced echo chambers in structurally polarized regimes where information spreads more easily. However, we also find echo chambers in regimes that do not correspond with polarized systems. This suggests that, contrary to what is commonly assumed, polarization and echo chambers are not equivalent concepts. The robustness of our results is confirmed on different network topologies, including two real-life datasets, a blog citation network \cite{polblogs} and a collection of U.S.\ congress bill co-sponsorship networks \cite{neal2020}. As a by-product, we also introduce a simple procedure to transform an unsigned network into a signed one with a prescribed degree of balance.

\section{Modeling} \label{modeling}
\subsection{Signed social networks} \label{signed_social_networks}

We consider an undirected signed network with $N$ agents. Each agent is assigned a latent group membership, which represents the divergent opinions within the population. If all in-group interactions are friendly and all out-group interactions are hostile, then the network is perfectly balanced. In the opposite case, where in-groups are always hostile towards each other and friendly with out-groups, we call the network perfectly {antibalanced} \cite{harary1957structural}. To interpolate between these two extremes, we employ the \textit{triadic degree of balance} $\tau$, also called \textit{triadic influence} \cite{ruiz2023triadic}. This index has been commonly used in the literature to quantify the level of (structural) polarization in a given social system \cite{aref2020multilevel, neal2020, gallo2024testing}. It is defined as:
\begin{equation}
	\bal = \text{tr}(W^3)/\text{tr}(\vert W\vert^3),
\end{equation}
where $W$ is the signed adjacency matrix of the network and $|W|$ its entrywise absolute value. In perfectly balanced networks $\tau=+1$, and in perfectly antibalanced ones $\tau=-1$. Values in-between quantify intermediate degrees of balance, with $\tau\approx0$ for networks that are neither significantly balanced nor antibalanced. The definition of $\tau$ is based on the product of link signs in triangles, which have been found to be an accurate and computationally fast proxy for balance \footnote{Triangles with a positive product of link signs famously respect the old adage ``the friend of my friend is my friend, the enemy of my friend if my enemy, the enemy of my enemy is my friend''.}. Hereafter, we employ the terms balanced and antibalanced to refer to networks with $\tau>0$ and $\tau<0$ respectively, while we reserve the terms perfectly (anti)balanced for the cases $\tau=\pm1$.

The majority of our simulations are run on networks with an underlying Erdös-Rényi (ER) topology with connection probability $p$. Hence, they do not exhibit communities from an unsigned point of view. Nevertheless, group memberships can induce mesoscopic structures affecting the degree of balance. We thus attribute signs to the links following a procedure that guarantees a pre-specified triadic degree of balance on average, inspired by \cite{cucuringu_2019}. To each link whose endpoints fall within the same group, we attribute a negative sign with probability $\eta$ and a positive sign otherwise; whereas links whose endpoints fall in different groups are positive with probability $\eta$ and negative otherwise. The  parameter $\eta$ regulates the degree of balance in the network. For $\eta=0$ the network is (perfectly) balanced, for $\eta=1/2$ link signs are uncorrelated with group memberships, while for $\eta=1$ we obtain an antibalanced network. Intermediate values allow us to interpolate between the different degrees of balance. 

\begin{figure}[!t]
    \centering
    \hspace{-.5cm}
    \includegraphics[width=.3\textwidth]{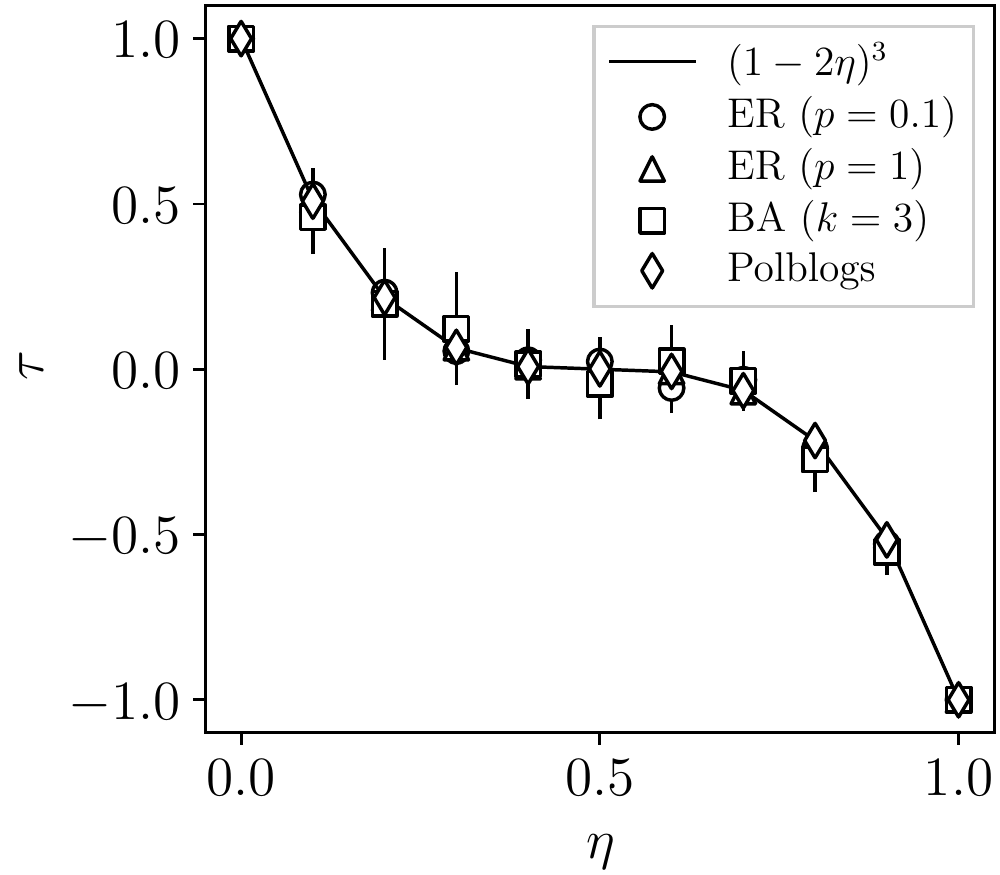}
	\caption{Degree of balance $\tau$ as a function of $\eta$, for various networks: ER with density $p=0.1$, ER with density $p=1.0$, BA with average degree $k=3$, and Polblogs \cite{polblogs} (details of the Polblogs dataset in Sec.~\ref{sec:robustness}). Each point is averaged over 10 random assignments of link signs. Vertical bars indicate one standard deviation from the mean.}
	\label{noise_balance}
\end{figure}

To generate networks with a specific triadic degree of balance $\tau$, we make use of the following nonlinear relation, proven in the Supplementary Material \cite[Sec.\ IIA]{supplementary}:
\begin{equation} \label{noise_bal_eq}
   { \tau} = (1-2\eta)^3
\end{equation} 
on expectation. Therefore, setting $\eta=(1-\bal^{1/3})/2$ will yield an ensemble of networks with average degree of balance $\bal$. This result is topology-independent and holds for any given network. We use it to control precisely the degree of balance of the networks we generate for our experiments. The relation is illustrated in Fig.~\ref{noise_balance} for networks of various topologies.

\subsection{Information propagation}
We assume that an item is propagating throughout the network: a journal article, a tweet, a picture, etc. To describe its propagation, we adapt the archetypal Independent Cascade Model of simple contagion \cite{kempe2003}. In this model, agents are initially all inactive (state 0), except for a few \emph{seeds} who are the original emitters of the item (state 1). Their neighbors are thus exposed to the item, and may choose to share it themselves. Formally, in consecutive steps, each agent that was activated in the previous step activates each of its neighbors individually with probability $\lambda$. This probability is called the activation probability. If all attempted activations in a single step fail, the process stops. To account for signed links, we add a novel feature to these dynamics and propose the Signed Independent Cascade Model (SICM). Active agents can now be in one of two states $\{-1,+1\}$, reflecting two opposite-minded ways to frame the information: a fresh government decision may spark praise from some and criticism from others. When an agent is activated through a positive (resp.\ negative) link, they adopt the same (resp.\ opposite) state of the agent responsible for their activation. Once an agent is active, its state is fixed and never changes. Technical details of the implementation can be found in the Supplementary Material \cite[Sec.\ IA]{supplementary}.

The SICM relies on a simple contagion mechanism: one exposure to information is enough to trigger a potential activation. In contrast, complex contagion mechanisms necessitate multiple exposures. This creates a different phenomenology \cite{Centola2007, Centola2018, min2018competition} that has been argued to better model realistic behavioral dynamics \cite{Centola2010, karsai2014complex, cencetti2023distinguishing}. For this purpose, we propose the Signed Linear Threshold Model (SLTM), adapted from \cite{kempe2003, watts2002simple}. Similarly as for the SICM, agents can be in one of two states $\{-1,+1\}$ depending on who triggered their activation. Agents activate if the aggregated influence of their neighbors exceeds a threshold $T$. Details of the modeling and implementation can be found in the Supplementary Material \cite[Sec.\ IB]{supplementary}. While most of our results are presented for the SICM, we show in Sec.~\ref{sec:robustness} that the SLTM exhibits similar behaviors.
 
\subsection{Echo chamber effect} 
To measure the intensity of the echo chamber effect, we adapt a metric from \cite{diazdiaz2022}. Let $\rho_{ab}^+$ denote the expected fraction of agents in state $+1$ among all the active agents in group $b$, at the end of a process with all seeds belonging to group $a$. Similarly, we define $\rho_{ab}^-,\rho_{ba}^+,\rho_{ba}^-$ as well as $\rho_{aa}^+,\rho_{aa}^-,\rho_{bb}^+,\rho_{bb}^-$, excluding the seeds from the computation of the last four. The echo chamber strength is defined as
\begin{equation} 
	E = \frac{\rho_{aa}^+ - \rho_{aa}^- - \rho_{ab}^+ + \rho_{ab}^- + \rho_{bb}^+ - \rho_{bb}^- -\rho_{ba}^+ + \rho_{ba}^-}{4}. \label{E_def}
\end{equation}
In the Supplementary Material \cite[Sec. IIB]{supplementary} we prove that $E$ is a second-order moment; thus, we can interpret $E$ as a measure of the correlation between states and group memberships. The maximum $E=+1$ corresponds to perfect echo chambers, where everyone inside a group is in the same state, opposite to that of the other group. The minimum $E=-1$ is the case of an \emph{anti-echo} chamber, where agents within the seed group disagree with the seeds, while agents of the opposite group agree with the seeds. When $E\approx0$, there is no significant echo chamber effect. In other words, $E$ measures the level of homogeneity in the framing of information within and between opposite-minded groups, rather than the presence or absence of cross-cutting exchanges. For the sake of conciseness, we refer to $E$ simply as the echo chamber effect, bearing in mind that it is actually a metric that quantifies this effect.

\section{Results} \label{results}
\subsection{Simple contagion on Erdös-Rényi networks}

\begin{figure}[t]
    \centering
    \includegraphics[width=.49\textwidth]{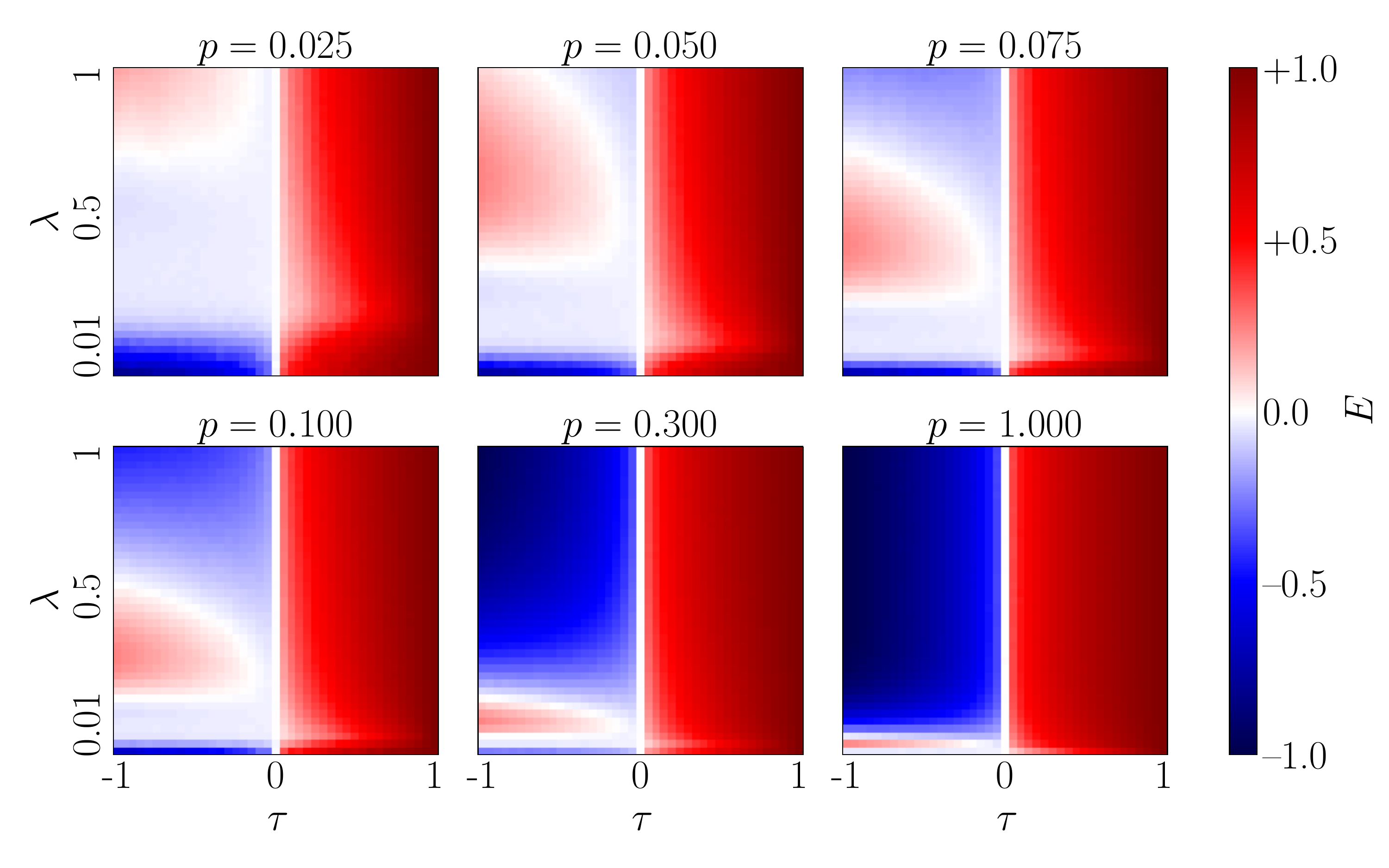}
    \caption{Echo chamber effect $E$ on signed Erdös-Rényi (ER) networks of various densities $p$, for SICM dynamics and as a function of the activation probability $\lambda$ and the triadic balance $\tau$. On the right half of the heatmaps are balanced networks $(\tau>0)$, on the left half are antibalanced networks $(\tau<0)$. In this figure and the remaining ones, we set $N=250$ nodes and $M=800$ realizations per data point.}
    \label{ssbm_simple}
\end{figure}

\begin{figure}[t]
    \begin{subfigure}[t]{.47\textwidth}
        \centering
        \includegraphics[width=\textwidth]{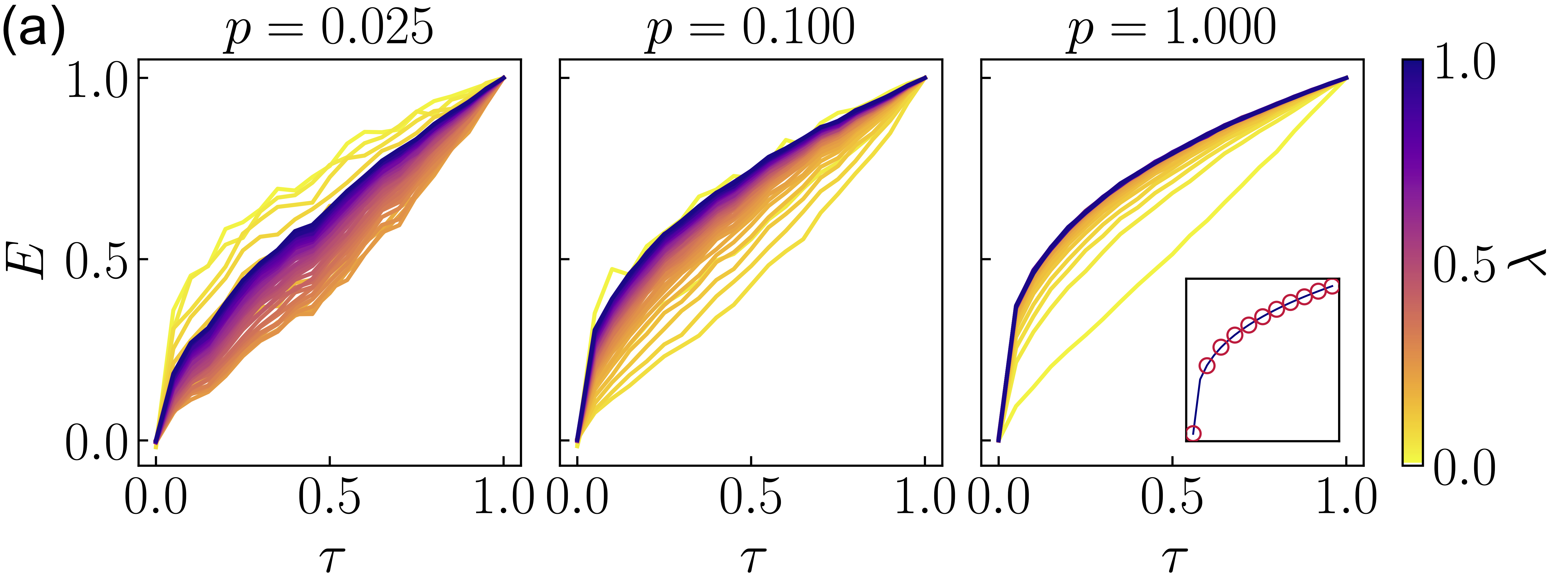}
    \end{subfigure}\\
    \begin{subfigure}[t]{.47\textwidth}
        \centering
        \includegraphics[width=\textwidth]{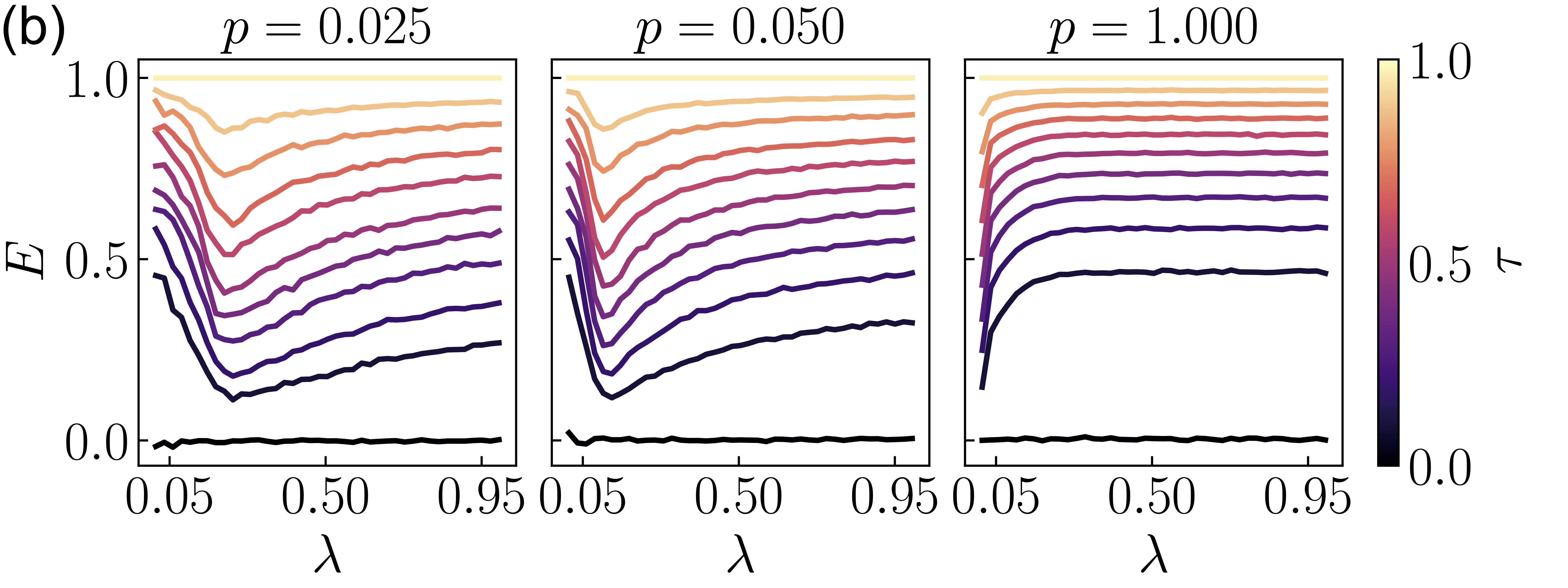}
    \end{subfigure}\\
    \begin{subfigure}[t]{.23\textwidth}
        \centering
        \includegraphics[width=\textwidth]{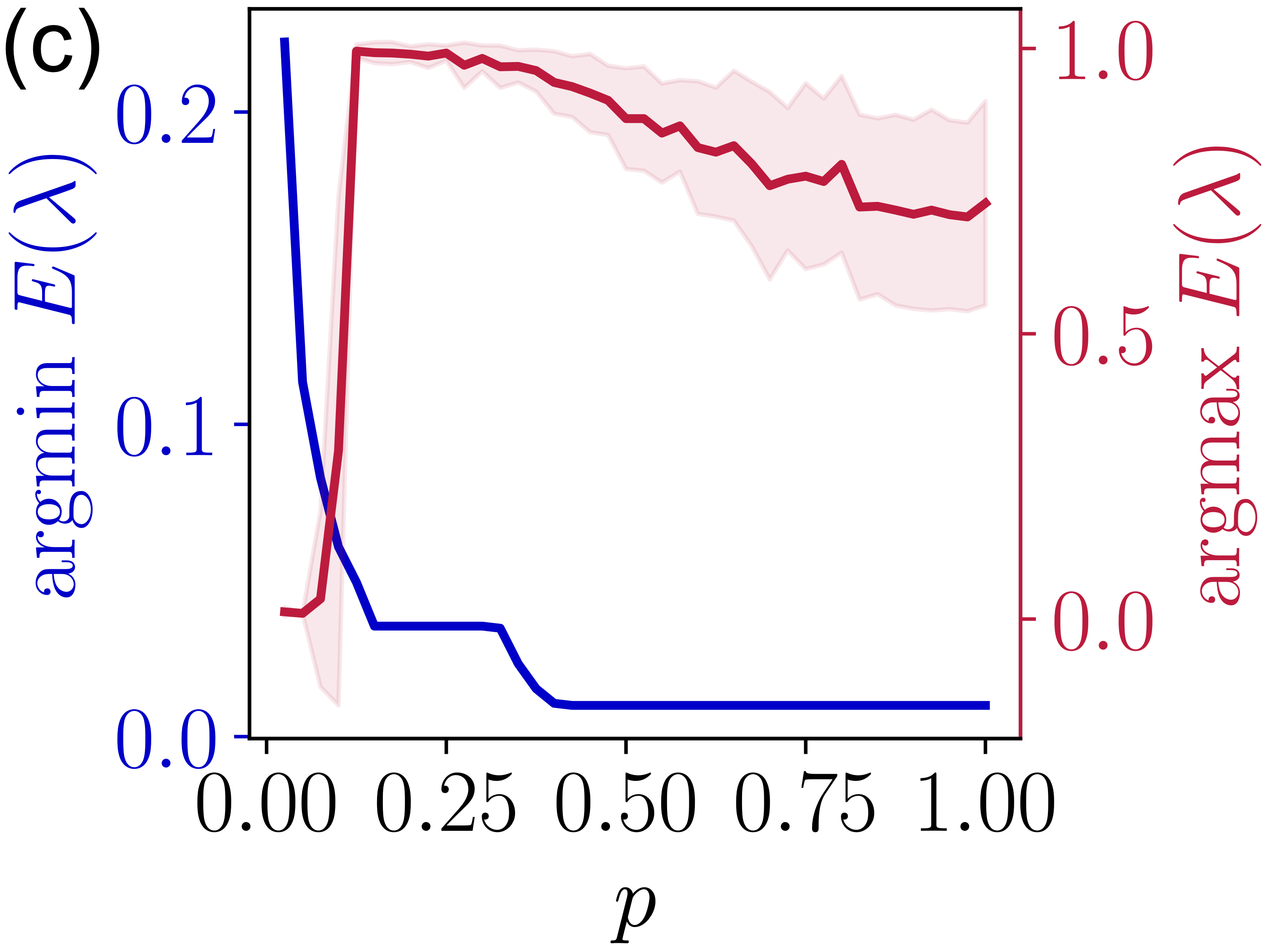}
    \end{subfigure}~
    \begin{subfigure}[t]{.23\textwidth}
        \centering
        \includegraphics[width=\textwidth]{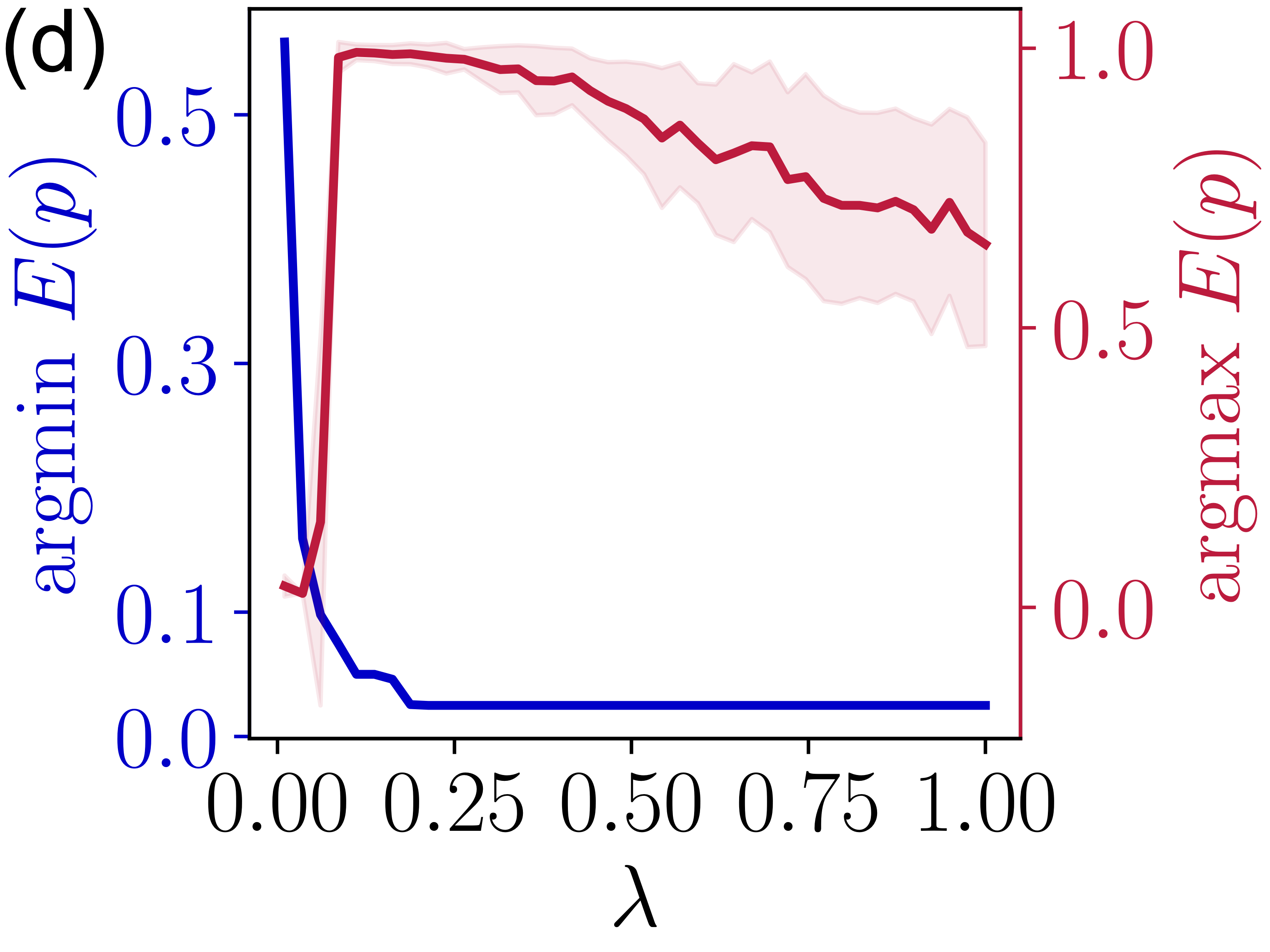}
    \end{subfigure}
    \caption{Echo chamber effect $E$ in balanced ER networks for SICM dynamics. (a) $E$ as a function of the triadic balance $\tau$ for various activation probabilities $\lambda$. The inset on the right plot shows the match between simulation results (red circles) and theoretical prediction ($E=\tau^{1/3}$, blue line) for $\lambda=p=1$. (b) $E$ as a function of $\lambda$ for various $\tau$. (c) activation probability realising the minimum and maximum echo chamber effect for various $p$, averaged over all values of $\tau$. (d) Density $p$ realising the minimum and maximum echo chamber effect for various $\lambda$, averaged over all values of $\tau$. Shaded areas cover one standard deviation from the mean.}
    \label{ssbm_pos}
\end{figure}

We assess the echo chamber effect via simulations of the SICM on ER networks with $N=250$ agents. Figure~\ref{ssbm_simple} shows $E$ as function of $\tau$ and $\lambda$ on networks of various densities $p$. The difference between balanced ($\tau>0$) and antibalanced ($\tau<0$) networks is immediately striking, with a transition between the two that gets sharper as $p$ increases. In balanced networks, $E$ increases monotonically with $\tau$, as shown in Fig.\ref{ssbm_pos}(a). Thus, we find that a stronger balance entails a higher echo chamber effect. In other words, echo chambers, much like polarization \cite{neal2020, talaga2023polarization} and dissensus \cite{altafini2013}, are amplified by the level of structural balance in balanced signed social networks. Note, however, that even in this simple scenario, the relation between the level in balance and the echo chamber effect is non-linear. In fact, in the Supplementary Material \cite[Sec. IIC]{supplementary}, we prove that for $\lambda=p=1$, this relation is given by $E=\tau^{1/3}$. See the inset in Fig.\ref{ssbm_pos}(a) for illustration.

In Fig.~\ref{ssbm_pos}(b), we notice that $E$ is also not monotonous with respect to $\lambda$. There exists a nontrivial value of $\lambda$ for which $E$ is minimal, which decreases continuously with the network density (Fig.~\ref{ssbm_pos}(c)). In contrast, the argmax of $E$ jumps discontinuously from $\lambda\approx 0$ to $\lambda\approx 1$. The same phenomenon is observed when computing the argmax of $E$ over $p$ (Fig. ~\ref{ssbm_pos}(d)). This suggests that minimizing the echo chamber effect is simpler than maximizing it: the optimal values of $\lambda$ and $p$ regarding minimization span a narrower range of values and are less susceptible to perturbations, whereas the optimal values for maximization vary widely between $0$ and $1$ and can change abruptly in the critical region. 

\begin{figure}[t]
    \begin{subfigure}[t]{.48\textwidth}
        \centering
        \includegraphics[width=\textwidth]{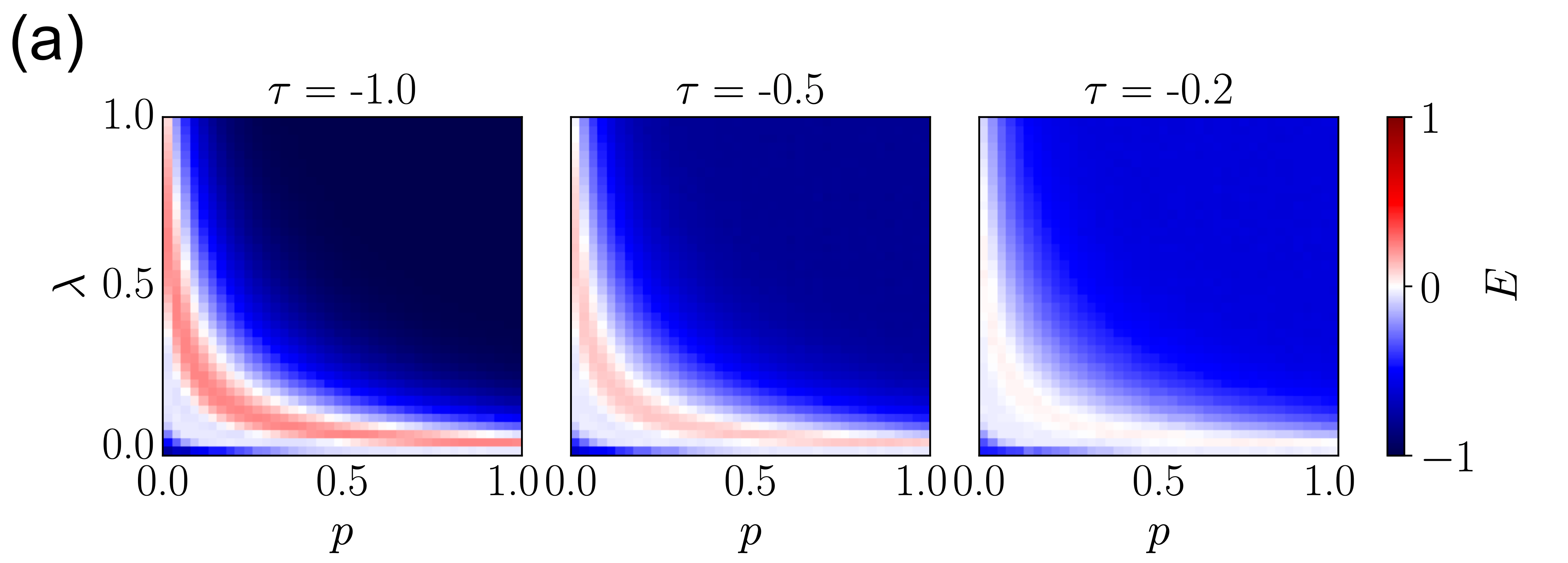}
    \end{subfigure}\\
    \begin{subfigure}[t]{.48\textwidth}
        \includegraphics[width=\textwidth]{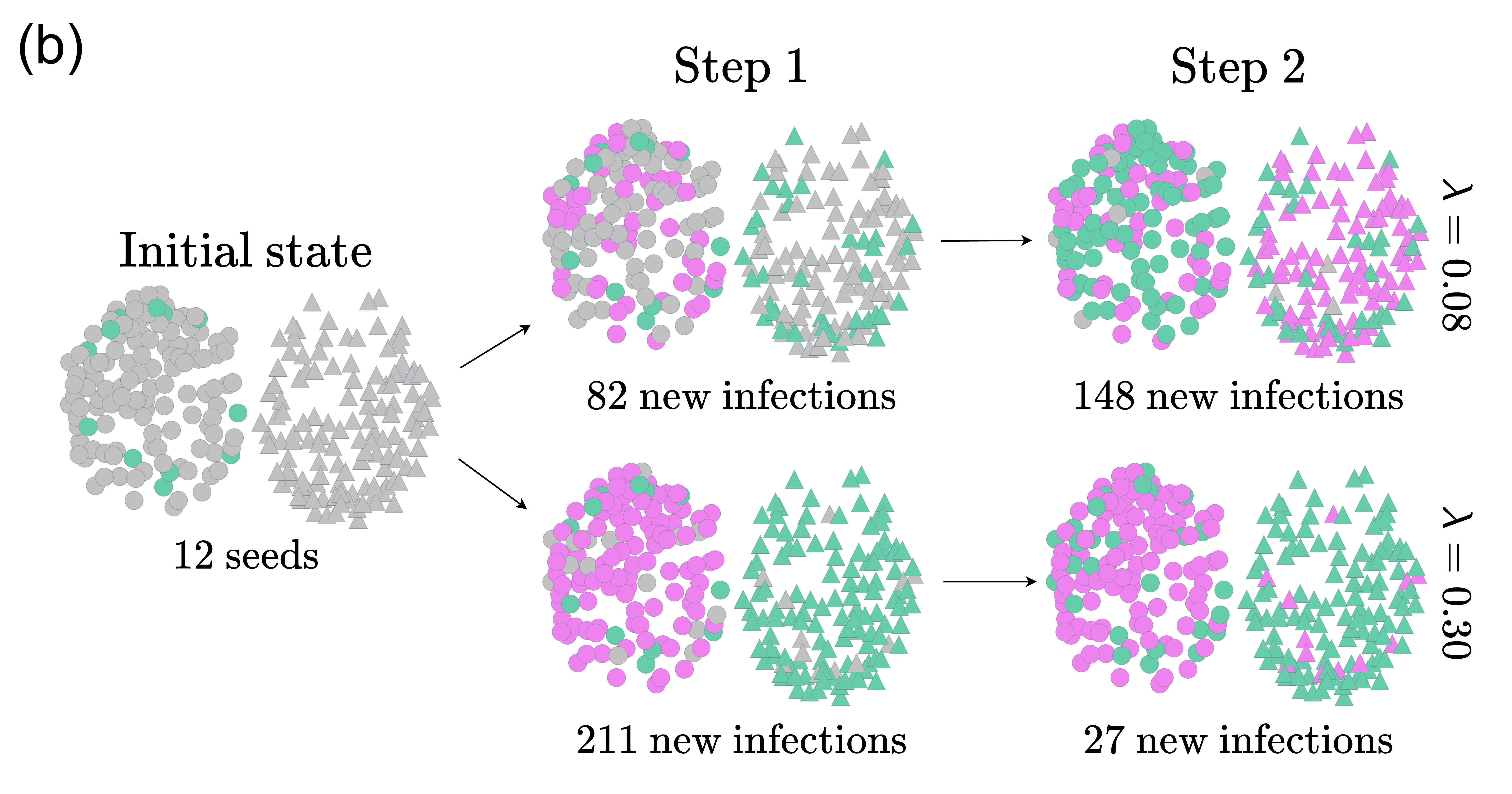}
    \end{subfigure}
    \caption{(a) Echo chamber effect as a function of network density and activation probability in antibalanced ER networks, for SICM dynamics and three different $\tau$. (b) Two realisations of the SICM on a signed ER network with $p=0.5,\tau=-1$, for $\lambda=0.08$ (echo chamber) and $\lambda=0.30$ (anti echo chamber). Circles correspond to agents of  group $a$ and triangles to agents of  group $b$. Agents are colored according to their state: grey for state $0$, green for state $+1$ and pink for state $-1$.}
    \label{ssbm_neg}
\end{figure}

In antibalanced networks, one could expect the same behavior, with $E$ decreasing monotonously towards $-1$ as $\tau$ becomes more negative. As we see in Fig.~\ref{ssbm_simple} however, this is not the case. Instead, we find that $E$ changes non-monotonously with both $p$ and $\lambda$ and $\tau$ (see the Supplementary Material \cite[Sec. III]{supplementary}). Specifically, $E$ reaches its minimum value, $E=-1$, both when $p$ and $\lambda$ are very high (e.g.\ $p=\lambda=1$) and when they are very low (e.g.\ $p=0.025, \lambda=0.01$). This phenomenon explicitly challenges the identification of structural polarization with echo chambers. Perhaps most surprisingly, certain  regions of the $(\tau<0,\lambda)$ half-plane display \textit{positive} echo chamber effects. This anomalous region, corresponding to the red semi-circular area of Fig. \ref{ssbm_simple}, shifts downwards as the network density $p$ increases, and almost vanishes at high densities. Interestingly, Figure~\ref{ssbm_neg}(a) reveals that for any given $\tau$, the boundaries of the anomalous region with respect to $p$ and $\lambda$ do not move. Instead, the value of $\tau$ only influences the strength of this anomaly.

To better understand this phenomenon, we compare how the propagation unfolds for two different values of the activation probability. Figure~\ref{ssbm_neg}(b) shows simulations of the SICM on a highly dense, perfectly antibalanced network ($\tau=-1$) with $\lambda=0.08$ and $\lambda=0.30$. Because $\tau=-1$, agents activated during odd steps adopt state $-1$ if they are in the seed group and state $+1$ otherwise, while the opposite happens during even steps. For $\lambda=0.30$, most activations occur in the first step, leading the seeding group to predominantly adopt state $-1$, resulting in a negative echo chamber effect. However, with $\lambda=0.08$, fewer activations happen initially, and most agents are activated in the second step. This causes the seeding group to adopt state $+1$, and generates a positive echo chamber effect even though the network is perfectly antibalanced. This alternance mechanism leads us to believe that this phenomenon is due to the synchronous aspect of the propagation process. Indeed, we show in the Supplementary Material \cite[Sec.\ III]{supplementary} that these anomalies disappear with asynchronous activations. Additionally, the exact shape of the anomalous regions seem to change slightly with the network size (Supplementary Material \cite[Sec.\ III]{supplementary}).

\subsection{Robustness} \label{sec:robustness}
\begin{figure}[t]
    \centering
    \includegraphics[width=.48\textwidth]{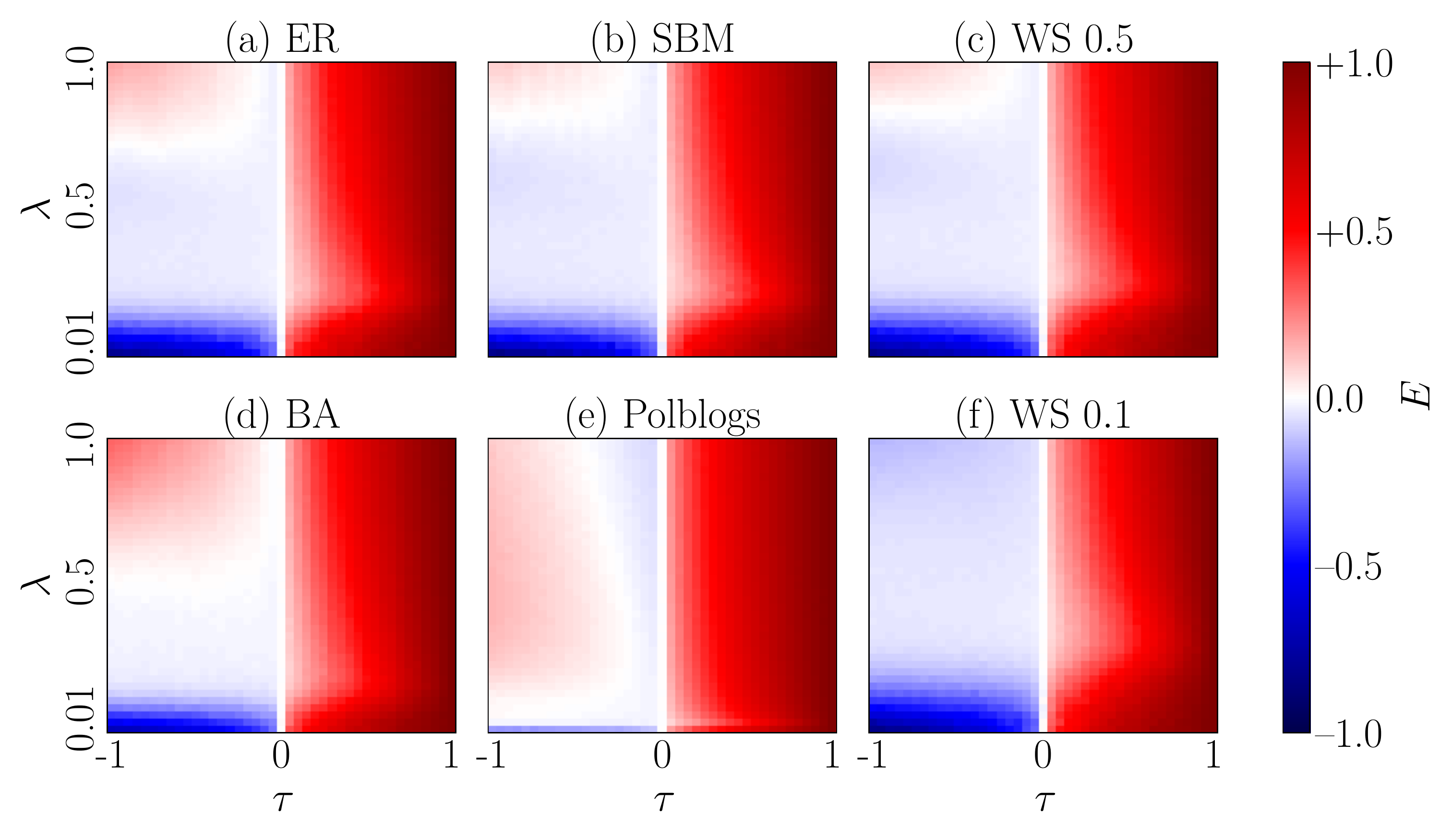}
    \caption{Echo chamber effect for SICM dynamics, with different networks topologies. (a) ER network with $p=0.025$. (b) Stochastic Block Model (SBM) with in-group link density $p_{in}=0.04$ and out-group link density $p_{out}=0.01$. Community memberships match  group memberships. (c) Watts-Strogatz (WS) network with average degree $k=6$ and rewiring probability $q=0.5$ (low clustering). (d) Barabási-Albert (BA) network with average degree $k=6$. (e) Polblogs network \cite{polblogs}. (f) WS network with average degree $k=6$ and rewiring probability $q=0.1$ (high clustering). All networks have and $N=250$ agents, except Polblogs with $N=1,222$.}
    \label{compare}
\end{figure}

To demonstrate the robustness of our approach, we show that similar results are achieved \emph{(i)} on networks with different topologies, and \emph{(ii)} with our complex contagion model. In Fig.~\ref{compare}, we apply the SICM on a collection of networks with similar densities but various topological features. The Stochastic Block Model (SBM) generates networks with community structure, the model of Watts-Strogatz (WS) generates networks with clustering, and the model of Barabasi-Albert (BA) generates scale-free networks. Additionally, we consider a network extracted from the Poblogs dataset \cite{polblogs}, which exhibits all of the above. The dataset comprises an unsigned network of political blogs collected during the campaign of the 2004 U.S.\ presidential elections. Blogs split into two groups, liberal and conservative, and there is an edge between two blogs if one referred to the other at least once. We restrict ourselves to the largest connected component of the network, with 1,222 nodes and 19,021 links. Link signs are generated following the procedure described in Sec.~\ref{signed_social_networks}, to obtain a collection of networks with varying degrees of balance.

\begin{figure}[t!]
    \centering
    \includegraphics[width=.48\textwidth]{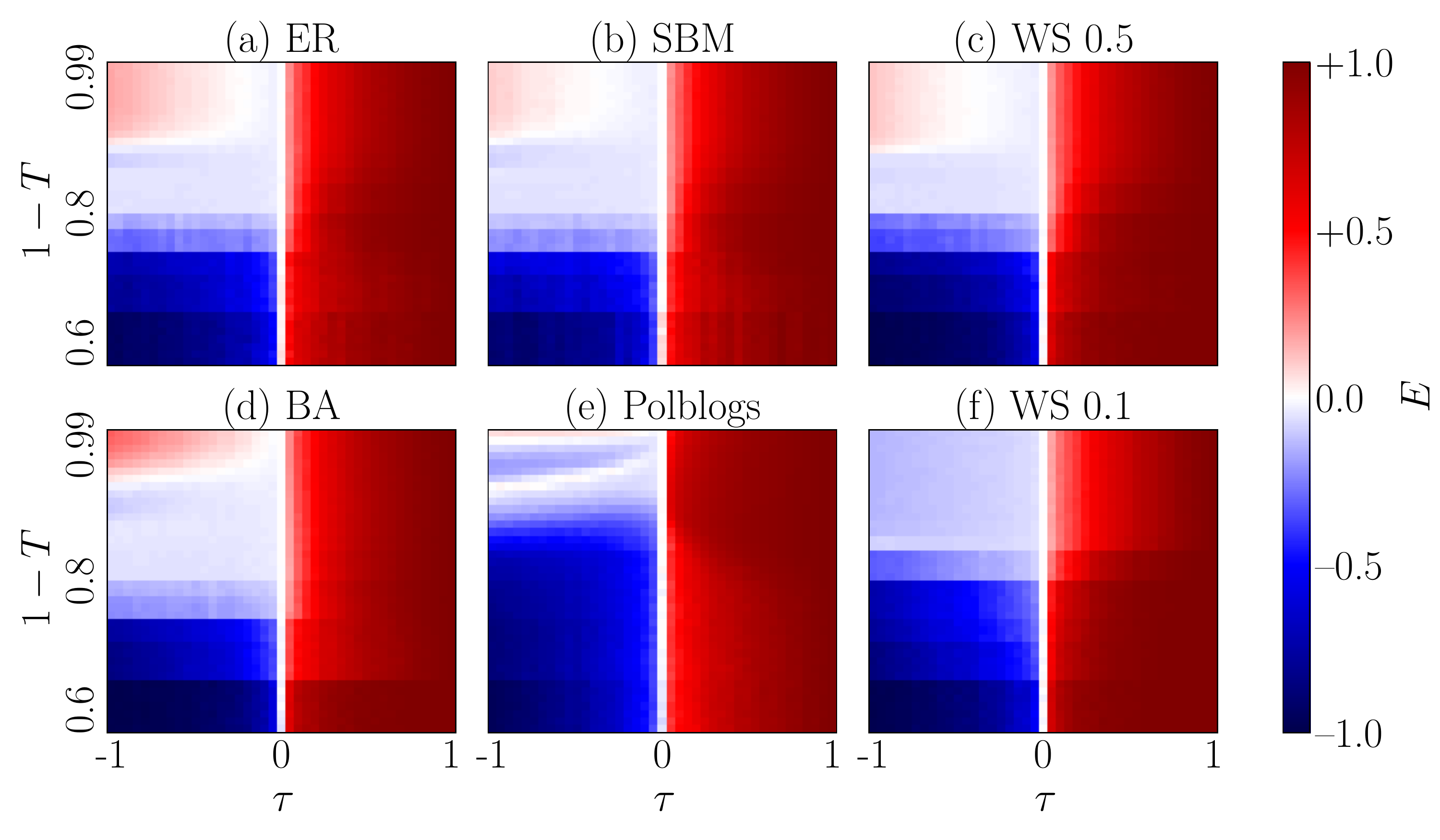}
    \caption{Echo chamber effect $E$ for a signed linear threshold model (SLTM) as a function of the triadic balance $\tau$ and the complementary of threshold parameter, $1-T$, across various topologies. The networks are the same as in Fig.~\ref{compare}.}
    \label{compare_complex}
\end{figure}

The echo chamber effect emerges when $\tau>0$ for every network, confirming that structural balance favors echo chambers regardless of the topology. Moreover, the anomalous region is also present for every network except the Watts-Strogatz one (Fig.~\ref{compare}f), due to clustering hindering the dominance of even-length paths. For complex contagion, we rely on the SLTM. Agents can now only activate if a sufficiently high number of their neighbors are active themselves. Even with this fundamental difference, our main results regarding the emergence of echo chambers hold. Figure~\ref{compare_complex} shows that SLTM dynamics again lead to a positive echo chamber effect when $\tau>0$, and an anomalous region in the $\tau<0$ half-plane. Slight differences with the SICM include sharper transitions, an increased presence of the anti-echo chamber effect in the $\tau<0$ half-plane, and the persistence of a few anomalous regions under asynchronous activations (Supplementary Material \cite[Sec.\ III]{supplementary}). Overall, these results suggest that the echo chamber phenomenology is robust with regards to the dynamical model employed.

\subsection{Real case study: U.S. congress}
Finally, we turn our attention to the study of a collection of real political signed networks between members of the U.S. congress \cite{neal2020}. Each network corresponds to a chamber, the House of Representatives (hereafter ``House'') or the Senate, and a congress number between 93\textsuperscript{rd} (1973-1975) and 114\textsuperscript{th} (2015-2017). Positive (negative) edges indicate that legislators co-sponsor significantly more (less) bills than expected under the corresponding null model.  The number of congress members varies between 443 and 452 for the House, between 100 and 109 for the Senate. Group memberships correspond to party affiliations, Democrat or Republican. There are a few independent congress members in the data, which we keep in the network but ignore when evaluating the echo chamber effect $E$. We refer the interested reader to the original article of Neal~\cite{neal2020} for more details. Importantly, in this network the signs of the edges are obtained from the empirical data, meaning that there is a predefined degree of balance $\tau$ for each network.

To investigate the echo chamber effect in this system, we simulate the SICM on all the networks with varying $\lambda$. The results are presented in Fig.~\ref{congress_EC}, which shows $E$ as a function of $\tau$. We obtain similar results with the SLTM, see the Supplementary Material \cite[Sec. III]{supplementary}. The trend is mostly increasing; however, there are three clear instances deviating from this correlation, marked with a blue circle and corresponding to the 93\textsuperscript{rd}, 94\textsuperscript{th} and 95\textsuperscript{th} congresses. This implies that the identification of polarization with echo chambers, while holding true for the majority of cases, can fail under certain structural patterns. To understand the origin of this discrepancy, we have analyzed four characteristics of the network as a function of time: the triadic level of balance $\tau$, the graph density, the ratio of negative to positive edges, and the ratio of Democrats to Republicans (Figure~\ref{congress_stats}). Clearly, the House of Representatives displays several unique structural features during the years where the discrepancy is observed. In particular, we find a very low density of edges and a small ratio of negative edges. We surmise that the combined effect of these two anomalies artificially increase the level of balance and thus the structural polarization, without necessarily impeding the communication between factions. As an aside, note that Fig. \ref{congress_stats}(a) also exhibits a negative degree of balance during the Carter and Reagan administrations, showing that antibalanced networks arise naturally in real-world systems.

\begin{figure}[t]
    \centering
    \includegraphics[width=.48\textwidth]{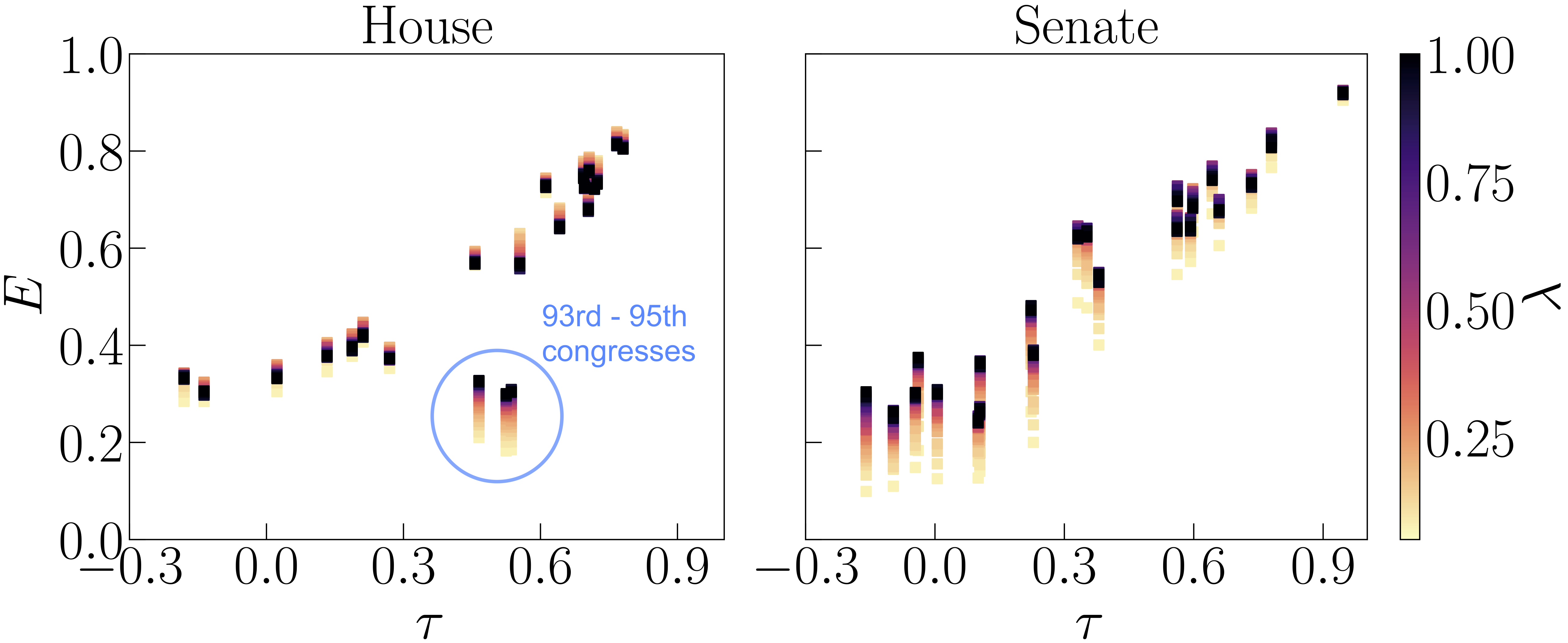}
    \caption{Echo chamber effect $E$ as a function of the degree of balance $\tau$, for SICM dynamics on the U.S.\ congress networks with various activation probabilities $\lambda$. We perform $M=400$ simulations per data point.}
    \label{congress_EC}
\end{figure}

\begin{figure}[t]
    \centering
    \includegraphics[width=.48\textwidth]{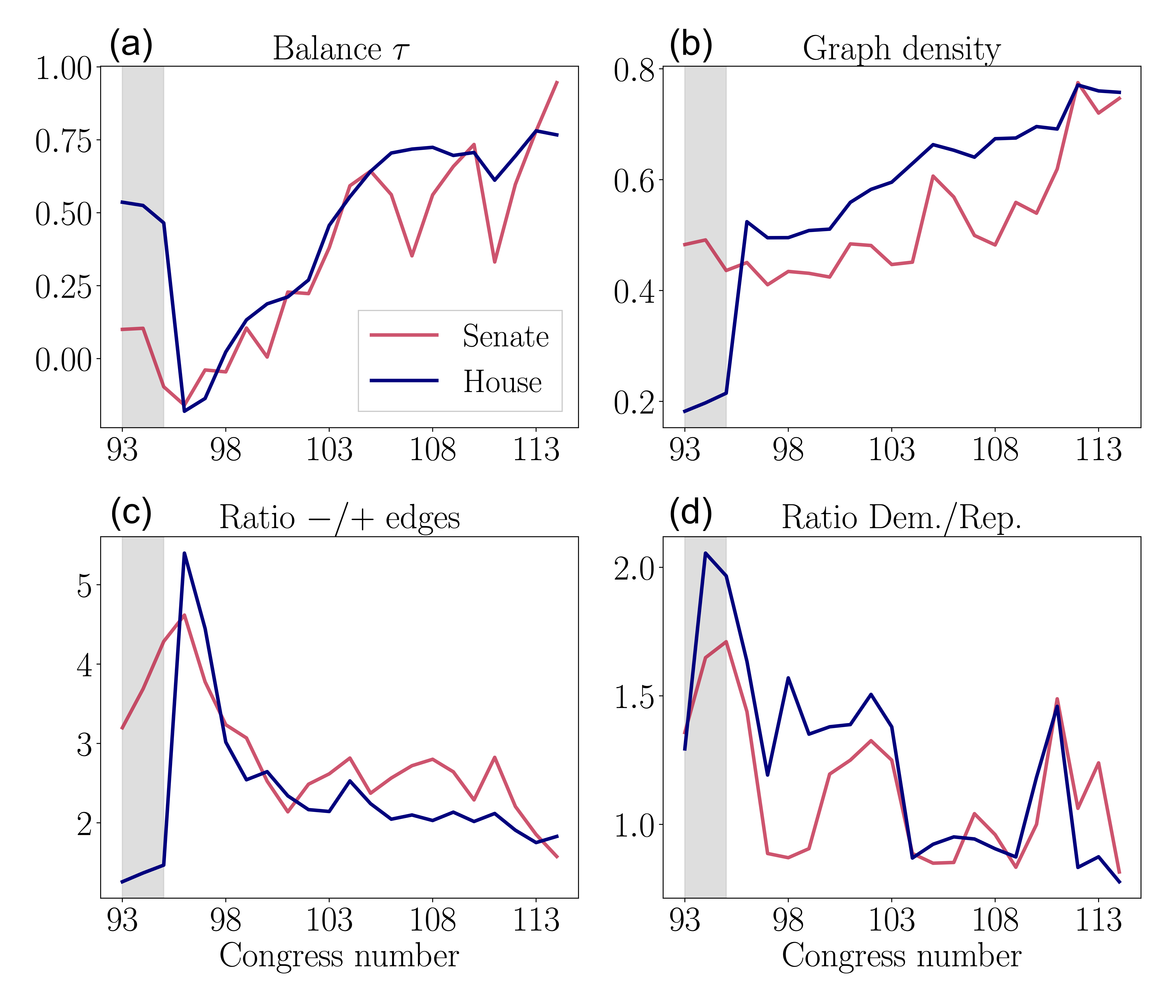}
    \caption{Evolution of the U.S.\ congress network metrics over time. (a) Degree of balance. (b) Network density. (c) Ratio of negative over positive edges. (d) Ratio of Democrats over Republicans. The grey vertical stripes cover the period from the 93\textsuperscript{rd} to the 96\textsuperscript{th} congresses, during which the echo chamber effect is unexpectedly low with respect to the degree of balance (cf.\ Fig.~\ref{congress_EC}).}
    \label{congress_stats}
\end{figure}

\section{Discussion} \label{discussion}
Our study makes two key contributions. First, we propose two novel contagion models, the SICM and the SLTM, that address how information can be framed differently among antagonistic groups. Second, and more importantly, we employ these models to describe the role of polarization, which we identify with structural balance, in the emergence of echo chambers. Our findings align with previous results connecting polarization with structural balance \cite{talaga2023polarization, neal2020, fraxanet2024unpacking} and connectivity \cite{pham2020}. However, while those studies focused on structural properties, ours is the first to emphasize the impact of these on information spreading. Importantly, we find cases where echo chambers are not correlated with polarization, highlighting that the connection between both phenomena is not always straightforward. Consequently, our findings point out to the dangers of identifying polarization with echo chambers. Our results hold for both simple and complex contagion and in a variety of network topologies, as well as in real-world networks. 

This work represents a first step towards incorporating the effect of negative links in the study of echo chambers. We highlighted that some of our results arise as the consequence of the choice of synchronous, discrete time steps for the model. Therefore, the relevance and consequences of different modelling choices should be further investigated. Future works could also consider the effects of structural balance beyond triangles \cite{aref2020multilevel, estrada2014walk}, other information transmission biases \cite{diazdiaz2022}, non-Markovian dynamics \cite{abella2023aging}, or homophily \cite{diazdiaz2022, perez2023polarized}. The theory of multilayer networks could also provide relevant insights into signed spreading phenomena \cite{kivela2014multilayer}. Moreover, our results could be applied to mitigate echo chamber effects and polarization in recommendation algorithms \cite{vendeville2022opening}. Finally, the SICM and SLTM show nontrivial regimes and transitions that could be explored more thoroughly. In this context, an analytic characterization of the contagion processes in signed networks would be highly valuable. Standard frameworks such as Mean Field Theory and the Approximate Master Equation \cite{gleeson2013binary} have been adapted to include signed edges in binary state dynamics \cite{krawiecki2024q,lee2023threshold,unicomb2018threshold}; however, to the best of our knowledge, no work has considered three-state dynamics with signed edges. We hope that our results motivate researchers to explore this new direction.

\begin{acknowledgments}
We want to thank the young researchers of the Complex Systems Society for financing a research visit of AV to IFISC through the Bridge Grants program, that served as the primary basis for this work. We thank Maxi San Miguel for insightful discussions and David Abella for suggesting improvements to the first draft. We also thank the anonymous reviewers for their precious remarks and suggestions that helped us improve this manuscript. FDD thanks financial support from MDM-2017-0711-20-2 funded by MCIN/AEI/10.13039/501100011033 and FSE, as well as project APASOS (No. PID2021-122256NB-C22). In the early stages of this research, AV was funded by the UK EPSRC grant EP/S022503/1 that supports the Centre for Doctoral Training in Cybersecurity delivered by UCL's Departments of Computer Science, Security and Crime Science, and Science, Technology, Engineering and Public Policy.
\end{acknowledgments}


\nocite{vendeville2024discord} 
\bibliographystyle{apsrev4-2}
\bibliography{biblio}

\providecommand{\noopsort}[1]{}\providecommand{\singleletter}[1]{#1}%
\begin{thebibliography}{72}%
\makeatletter
\providecommand \@ifxundefined [1]{%
 \@ifx{#1\undefined}
}%
\providecommand \@ifnum [1]{%
 \ifnum #1\expandafter \@firstoftwo
 \else \expandafter \@secondoftwo
 \fi
}%
\providecommand \@ifx [1]{%
 \ifx #1\expandafter \@firstoftwo
 \else \expandafter \@secondoftwo
 \fi
}%
\providecommand \natexlab [1]{#1}%
\providecommand \enquote  [1]{``#1''}%
\providecommand \bibnamefont  [1]{#1}%
\providecommand \bibfnamefont [1]{#1}%
\providecommand \citenamefont [1]{#1}%
\providecommand \href@noop [0]{\@secondoftwo}%
\providecommand \href [0]{\begingroup \@sanitize@url \@href}%
\providecommand \@href[1]{\@@startlink{#1}\@@href}%
\providecommand \@@href[1]{\endgroup#1\@@endlink}%
\providecommand \@sanitize@url [0]{\catcode `\\12\catcode `\$12\catcode
  `\&12\catcode `\#12\catcode `\^12\catcode `\_12\catcode `\%12\relax}%
\providecommand \@@startlink[1]{}%
\providecommand \@@endlink[0]{}%
\providecommand \url  [0]{\begingroup\@sanitize@url \@url }%
\providecommand \@url [1]{\endgroup\@href {#1}{\urlprefix }}%
\providecommand \urlprefix  [0]{URL }%
\providecommand \Eprint [0]{\href }%
\providecommand \doibase [0]{https://doi.org/}%
\providecommand \selectlanguage [0]{\@gobble}%
\providecommand \bibinfo  [0]{\@secondoftwo}%
\providecommand \bibfield  [0]{\@secondoftwo}%
\providecommand \translation [1]{[#1]}%
\providecommand \BibitemOpen [0]{}%
\providecommand \bibitemStop [0]{}%
\providecommand \bibitemNoStop [0]{.\EOS\space}%
\providecommand \EOS [0]{\spacefactor3000\relax}%
\providecommand \BibitemShut  [1]{\csname bibitem#1\endcsname}%
\let\auto@bib@innerbib\@empty
\bibitem [{\citenamefont {Baumann}\ \emph {et~al.}(2020)\citenamefont
  {Baumann}, \citenamefont {Lorenz-Spreen}, \citenamefont {Sokolov},\ and\
  \citenamefont {Starnini}}]{baumann2020}%
  \BibitemOpen
  \bibfield  {author} {\bibinfo {author} {\bibfnamefont {F.}~\bibnamefont
  {Baumann}}, \bibinfo {author} {\bibfnamefont {P.}~\bibnamefont
  {Lorenz-Spreen}}, \bibinfo {author} {\bibfnamefont {I.~M.}\ \bibnamefont
  {Sokolov}},\ and\ \bibinfo {author} {\bibfnamefont {M.}~\bibnamefont
  {Starnini}},\ }\href {https://doi.org/10.1103/PhysRevLett.124.048301}
  {\bibfield  {journal} {\bibinfo  {journal} {Phys. Rev. Lett.}\ }\textbf
  {\bibinfo {volume} {124}},\ \bibinfo {pages} {048301} (\bibinfo {year}
  {2020})}\BibitemShut {NoStop}%
\bibitem [{\citenamefont {Baqir}\ \emph {et~al.}(2023)\citenamefont {Baqir},
  \citenamefont {Chen}, \citenamefont {Diaz-Diaz}, \citenamefont {Kiyak},
  \citenamefont {Louf}, \citenamefont {Morini}, \citenamefont {Pansanella},
  \citenamefont {Torricelli},\ and\ \citenamefont
  {Galeazzi}}]{baqir2023beyond}%
  \BibitemOpen
  \bibfield  {author} {\bibinfo {author} {\bibfnamefont {A.}~\bibnamefont
  {Baqir}}, \bibinfo {author} {\bibfnamefont {Y.}~\bibnamefont {Chen}},
  \bibinfo {author} {\bibfnamefont {F.}~\bibnamefont {Diaz-Diaz}}, \bibinfo
  {author} {\bibfnamefont {S.}~\bibnamefont {Kiyak}}, \bibinfo {author}
  {\bibfnamefont {T.}~\bibnamefont {Louf}}, \bibinfo {author} {\bibfnamefont
  {V.}~\bibnamefont {Morini}}, \bibinfo {author} {\bibfnamefont
  {V.}~\bibnamefont {Pansanella}}, \bibinfo {author} {\bibfnamefont
  {M.}~\bibnamefont {Torricelli}},\ and\ \bibinfo {author} {\bibfnamefont
  {A.}~\bibnamefont {Galeazzi}},\ }\href@noop {} {\bibinfo {title} {Beyond
  active engagement: {The} significance of lurkers in a polarized {Twitter}
  debate}} (\bibinfo {year} {2023}),\ \Eprint
  {https://arxiv.org/abs/2306.17538} {arXiv:2306.17538} \BibitemShut {NoStop}%
\bibitem [{\citenamefont {Conover}\ \emph {et~al.}(2021)\citenamefont
  {Conover}, \citenamefont {Ratkiewicz}, \citenamefont {Francisco},
  \citenamefont {Goncalves}, \citenamefont {Menczer},\ and\ \citenamefont
  {Flammini}}]{conover2011}%
  \BibitemOpen
  \bibfield  {author} {\bibinfo {author} {\bibfnamefont {M.}~\bibnamefont
  {Conover}}, \bibinfo {author} {\bibfnamefont {J.}~\bibnamefont {Ratkiewicz}},
  \bibinfo {author} {\bibfnamefont {M.}~\bibnamefont {Francisco}}, \bibinfo
  {author} {\bibfnamefont {B.}~\bibnamefont {Goncalves}}, \bibinfo {author}
  {\bibfnamefont {F.}~\bibnamefont {Menczer}},\ and\ \bibinfo {author}
  {\bibfnamefont {A.}~\bibnamefont {Flammini}},\ }in\ \href
  {https://doi.org/10.1609/icwsm.v5i1.14126} {\emph {\bibinfo {booktitle}
  {Proceedings of the International AAAI Conference on Web and Social
  Media}}},\ Vol.~\bibinfo {volume} {5}\ (\bibinfo {year} {2021})\ pp.\
  \bibinfo {pages} {89--96}\BibitemShut {NoStop}%
\bibitem [{\citenamefont {Halberstam}\ and\ \citenamefont
  {Knight}(2016)}]{halberstam2016}%
  \BibitemOpen
  \bibfield  {author} {\bibinfo {author} {\bibfnamefont {Y.}~\bibnamefont
  {Halberstam}}\ and\ \bibinfo {author} {\bibfnamefont {B.}~\bibnamefont
  {Knight}},\ }\href {https://doi.org/10.1016/j.jpubeco.2016.08.011} {\bibfield
   {journal} {\bibinfo  {journal} {J. Public Econ.}\ }\textbf {\bibinfo
  {volume} {143}},\ \bibinfo {pages} {73} (\bibinfo {year} {2016})}\BibitemShut
  {NoStop}%
\bibitem [{\citenamefont {Garimella}\ \emph {et~al.}(2018)\citenamefont
  {Garimella}, \citenamefont {De~Francisci~Morales}, \citenamefont {Gionis},\
  and\ \citenamefont {Mathioudakis}}]{garimella2018}%
  \BibitemOpen
  \bibfield  {author} {\bibinfo {author} {\bibfnamefont {K.}~\bibnamefont
  {Garimella}}, \bibinfo {author} {\bibfnamefont {G.}~\bibnamefont
  {De~Francisci~Morales}}, \bibinfo {author} {\bibfnamefont {A.}~\bibnamefont
  {Gionis}},\ and\ \bibinfo {author} {\bibfnamefont {M.}~\bibnamefont
  {Mathioudakis}},\ }in\ \href {https://doi.org/10.1145/3178876.3186139} {\emph
  {\bibinfo {booktitle} {Proceedings of the 2018 World Wide Web Conference}}},\
  \bibinfo {series and number} {WWW '18}\ (\bibinfo  {publisher} {International
  World Wide Web Conferences Steering Committee},\ \bibinfo {address} {Republic
  and Canton of Geneva, CHE},\ \bibinfo {year} {2018})\ pp.\ \bibinfo {pages}
  {913--922}\BibitemShut {NoStop}%
\bibitem [{\citenamefont {Cinelli}\ \emph {et~al.}(2021)\citenamefont
  {Cinelli}, \citenamefont {De~Francisci~Morales}, \citenamefont {Galeazzi},
  \citenamefont {Quattrociocchi},\ and\ \citenamefont
  {Starnini}}]{cinelli2021}%
  \BibitemOpen
  \bibfield  {author} {\bibinfo {author} {\bibfnamefont {M.}~\bibnamefont
  {Cinelli}}, \bibinfo {author} {\bibfnamefont {G.}~\bibnamefont
  {De~Francisci~Morales}}, \bibinfo {author} {\bibfnamefont {A.}~\bibnamefont
  {Galeazzi}}, \bibinfo {author} {\bibfnamefont {W.}~\bibnamefont
  {Quattrociocchi}},\ and\ \bibinfo {author} {\bibfnamefont {M.}~\bibnamefont
  {Starnini}},\ }\href {https://doi.org/10.1073/pnas.2023301118} {\bibfield
  {journal} {\bibinfo  {journal} {Proc. Natl. Acad. Sci.}\ }\textbf {\bibinfo
  {volume} {118}},\ \bibinfo {pages} {e2023301118} (\bibinfo {year}
  {2021})}\BibitemShut {NoStop}%
\bibitem [{\citenamefont {Bakshy}\ \emph {et~al.}(2015)\citenamefont {Bakshy},
  \citenamefont {Messing},\ and\ \citenamefont {Adamic}}]{bakshy2015}%
  \BibitemOpen
  \bibfield  {author} {\bibinfo {author} {\bibfnamefont {E.}~\bibnamefont
  {Bakshy}}, \bibinfo {author} {\bibfnamefont {S.}~\bibnamefont {Messing}},\
  and\ \bibinfo {author} {\bibfnamefont {L.~A.}\ \bibnamefont {Adamic}},\
  }\href {https://doi.org/10.1126/science.aaa1160} {\bibfield  {journal}
  {\bibinfo  {journal} {Science}\ }\textbf {\bibinfo {volume} {348}},\ \bibinfo
  {pages} {1130} (\bibinfo {year} {2015})}\BibitemShut {NoStop}%
\bibitem [{\citenamefont {González-Bailón}\ \emph {et~al.}(2023)\citenamefont
  {González-Bailón}, \citenamefont {Lazer}, \citenamefont {Barberá},
  \citenamefont {Zhang}, \citenamefont {Allcott}, \citenamefont {Brown},
  \citenamefont {Crespo-Tenorio}, \citenamefont {Freelon}, \citenamefont
  {Gentzkow}, \citenamefont {Guess}, \citenamefont {Iyengar}, \citenamefont
  {Kim}, \citenamefont {Malhotra}, \citenamefont {Moehler}, \citenamefont
  {Nyhan}, \citenamefont {Pan}, \citenamefont {Rivera}, \citenamefont {Settle},
  \citenamefont {Thorson}, \citenamefont {Tromble}, \citenamefont {Wilkins},
  \citenamefont {Wojcieszak}, \citenamefont {de~Jonge}, \citenamefont {Franco},
  \citenamefont {Mason}, \citenamefont {Stroud},\ and\ \citenamefont
  {Tucker}}]{bailon2023}%
  \BibitemOpen
  \bibfield  {author} {\bibinfo {author} {\bibfnamefont {S.}~\bibnamefont
  {González-Bailón}}, \bibinfo {author} {\bibfnamefont {D.}~\bibnamefont
  {Lazer}}, \bibinfo {author} {\bibfnamefont {P.}~\bibnamefont {Barberá}},
  \bibinfo {author} {\bibfnamefont {M.}~\bibnamefont {Zhang}}, \bibinfo
  {author} {\bibfnamefont {H.}~\bibnamefont {Allcott}}, \bibinfo {author}
  {\bibfnamefont {T.}~\bibnamefont {Brown}}, \bibinfo {author} {\bibfnamefont
  {A.}~\bibnamefont {Crespo-Tenorio}}, \bibinfo {author} {\bibfnamefont
  {D.}~\bibnamefont {Freelon}}, \bibinfo {author} {\bibfnamefont
  {M.}~\bibnamefont {Gentzkow}}, \bibinfo {author} {\bibfnamefont {A.~M.}\
  \bibnamefont {Guess}}, \bibinfo {author} {\bibfnamefont {S.}~\bibnamefont
  {Iyengar}}, \bibinfo {author} {\bibfnamefont {Y.~M.}\ \bibnamefont {Kim}},
  \bibinfo {author} {\bibfnamefont {N.}~\bibnamefont {Malhotra}}, \bibinfo
  {author} {\bibfnamefont {D.}~\bibnamefont {Moehler}}, \bibinfo {author}
  {\bibfnamefont {B.}~\bibnamefont {Nyhan}}, \bibinfo {author} {\bibfnamefont
  {J.}~\bibnamefont {Pan}}, \bibinfo {author} {\bibfnamefont {C.~V.}\
  \bibnamefont {Rivera}}, \bibinfo {author} {\bibfnamefont {J.}~\bibnamefont
  {Settle}}, \bibinfo {author} {\bibfnamefont {E.}~\bibnamefont {Thorson}},
  \bibinfo {author} {\bibfnamefont {R.}~\bibnamefont {Tromble}}, \bibinfo
  {author} {\bibfnamefont {A.}~\bibnamefont {Wilkins}}, \bibinfo {author}
  {\bibfnamefont {M.}~\bibnamefont {Wojcieszak}}, \bibinfo {author}
  {\bibfnamefont {C.~K.}\ \bibnamefont {de~Jonge}}, \bibinfo {author}
  {\bibfnamefont {A.}~\bibnamefont {Franco}}, \bibinfo {author} {\bibfnamefont
  {W.}~\bibnamefont {Mason}}, \bibinfo {author} {\bibfnamefont {N.~J.}\
  \bibnamefont {Stroud}},\ and\ \bibinfo {author} {\bibfnamefont {J.~A.}\
  \bibnamefont {Tucker}},\ }\href {https://doi.org/10.1126/science.ade7138}
  {\bibfield  {journal} {\bibinfo  {journal} {Science}\ }\textbf {\bibinfo
  {volume} {381}},\ \bibinfo {pages} {392} (\bibinfo {year}
  {2023})}\BibitemShut {NoStop}%
\bibitem [{\citenamefont {Flaxman}\ \emph {et~al.}(2016)\citenamefont
  {Flaxman}, \citenamefont {Goel},\ and\ \citenamefont {Rao}}]{flaxman2016}%
  \BibitemOpen
  \bibfield  {author} {\bibinfo {author} {\bibfnamefont {S.}~\bibnamefont
  {Flaxman}}, \bibinfo {author} {\bibfnamefont {S.}~\bibnamefont {Goel}},\ and\
  \bibinfo {author} {\bibfnamefont {J.~M.}\ \bibnamefont {Rao}},\ }\href
  {https://doi.org/10.1093/poq/nfw006} {\bibfield  {journal} {\bibinfo
  {journal} {Public Opin. Q.}\ }\textbf {\bibinfo {volume} {80}},\ \bibinfo
  {pages} {298} (\bibinfo {year} {2016})}\BibitemShut {NoStop}%
\bibitem [{\citenamefont {Dubois}\ and\ \citenamefont
  {Blank}(2018)}]{dubois2018}%
  \BibitemOpen
  \bibfield  {author} {\bibinfo {author} {\bibfnamefont {E.}~\bibnamefont
  {Dubois}}\ and\ \bibinfo {author} {\bibfnamefont {G.}~\bibnamefont {Blank}},\
  }\href {https://doi.org/10.1080/1369118X.2018.1428656} {\bibfield  {journal}
  {\bibinfo  {journal} {Inf. Commun. Soc.}\ }\textbf {\bibinfo {volume} {21}},\
  \bibinfo {pages} {729} (\bibinfo {year} {2018})}\BibitemShut {NoStop}%
\bibitem [{\citenamefont {TianYang}\ \emph {et~al.}(2020)\citenamefont
  {TianYang}, \citenamefont {Majó-Vázquez}, \citenamefont {Nielsen},\ and\
  \citenamefont {González-Bailón}}]{yang2020}%
  \BibitemOpen
  \bibfield  {author} {\bibinfo {author} {\bibnamefont {TianYang}}, \bibinfo
  {author} {\bibfnamefont {S.}~\bibnamefont {Majó-Vázquez}}, \bibinfo
  {author} {\bibfnamefont {R.~K.}\ \bibnamefont {Nielsen}},\ and\ \bibinfo
  {author} {\bibfnamefont {S.}~\bibnamefont {González-Bailón}},\ }\href
  {https://doi.org/10.1073/pnas.2006089117} {\bibfield  {journal} {\bibinfo
  {journal} {Proc. Natl. Acad. Sci.}\ }\textbf {\bibinfo {volume} {117}},\
  \bibinfo {pages} {28678} (\bibinfo {year} {2020})}\BibitemShut {NoStop}%
\bibitem [{\citenamefont {De~Francisci~Morales}\ \emph
  {et~al.}(2021)\citenamefont {De~Francisci~Morales}, \citenamefont {Monti},\
  and\ \citenamefont {Starnini}}]{defranciscimorales2021}%
  \BibitemOpen
  \bibfield  {author} {\bibinfo {author} {\bibfnamefont {G.}~\bibnamefont
  {De~Francisci~Morales}}, \bibinfo {author} {\bibfnamefont {C.}~\bibnamefont
  {Monti}},\ and\ \bibinfo {author} {\bibfnamefont {M.}~\bibnamefont
  {Starnini}},\ }\href {https://doi.org/10.1038/s41598-021-81531-x} {\bibfield
  {journal} {\bibinfo  {journal} {Sci. Rep.}\ }\textbf {\bibinfo {volume}
  {11}},\ \bibinfo {pages} {2818} (\bibinfo {year} {2021})}\BibitemShut
  {NoStop}%
\bibitem [{\citenamefont {Guess}(2021)}]{guess2021}%
  \BibitemOpen
  \bibfield  {author} {\bibinfo {author} {\bibfnamefont {A.~M.}\ \bibnamefont
  {Guess}},\ }\href {https://doi.org/https://doi.org/10.1111/ajps.12589}
  {\bibfield  {journal} {\bibinfo  {journal} {Am. J. Pol. Sci.}\ }\textbf
  {\bibinfo {volume} {65}},\ \bibinfo {pages} {1007} (\bibinfo {year}
  {2021})}\BibitemShut {NoStop}%
\bibitem [{\citenamefont {Törnberg}(2022)}]{tornberg2022}%
  \BibitemOpen
  \bibfield  {author} {\bibinfo {author} {\bibfnamefont {P.}~\bibnamefont
  {Törnberg}},\ }\href {https://doi.org/10.1073/pnas.2207159119} {\bibfield
  {journal} {\bibinfo  {journal} {Proc. Natl. Acad. Sci.}\ }\textbf {\bibinfo
  {volume} {119}},\ \bibinfo {pages} {e2207159119} (\bibinfo {year}
  {2022})}\BibitemShut {NoStop}%
\bibitem [{\citenamefont {Williams}\ \emph {et~al.}(2015)\citenamefont
  {Williams}, \citenamefont {McMurray}, \citenamefont {Kurz},\ and\
  \citenamefont {{Hugo Lambert}}}]{williams2015}%
  \BibitemOpen
  \bibfield  {author} {\bibinfo {author} {\bibfnamefont {H.~T.}\ \bibnamefont
  {Williams}}, \bibinfo {author} {\bibfnamefont {J.~R.}\ \bibnamefont
  {McMurray}}, \bibinfo {author} {\bibfnamefont {T.}~\bibnamefont {Kurz}},\
  and\ \bibinfo {author} {\bibfnamefont {F.}~\bibnamefont {{Hugo Lambert}}},\
  }\href {https://doi.org/https://doi.org/10.1016/j.gloenvcha.2015.03.006}
  {\bibfield  {journal} {\bibinfo  {journal} {Glob. Environ. Change.}\ }\textbf
  {\bibinfo {volume} {32}},\ \bibinfo {pages} {126 } (\bibinfo {year}
  {2015})}\BibitemShut {NoStop}%
\bibitem [{\citenamefont {Tacchi}\ \emph {et~al.}(2022)\citenamefont {Tacchi},
  \citenamefont {Boldrini}, \citenamefont {Passarella},\ and\ \citenamefont
  {Conti}}]{tacchi2022}%
  \BibitemOpen
  \bibfield  {author} {\bibinfo {author} {\bibfnamefont {J.}~\bibnamefont
  {Tacchi}}, \bibinfo {author} {\bibfnamefont {C.}~\bibnamefont {Boldrini}},
  \bibinfo {author} {\bibfnamefont {A.}~\bibnamefont {Passarella}},\ and\
  \bibinfo {author} {\bibfnamefont {M.}~\bibnamefont {Conti}},\ }in\ \href
  {https://doi.org/10.1109/BigData55660.2022.10020939} {\emph {\bibinfo
  {booktitle} {Proceedings of the 2022 IEEE International Conference on Big
  Data}}}\ (\bibinfo {year} {2022})\ pp.\ \bibinfo {pages}
  {6030--6038}\BibitemShut {NoStop}%
\bibitem [{\citenamefont {Efstratiou}\ \emph {et~al.}(2023)\citenamefont
  {Efstratiou}, \citenamefont {Blackburn}, \citenamefont {Caulfield},
  \citenamefont {Stringhini}, \citenamefont {Zannettou},\ and\ \citenamefont
  {De~Cristofaro}}]{efstratiou2023}%
  \BibitemOpen
  \bibfield  {author} {\bibinfo {author} {\bibfnamefont {A.}~\bibnamefont
  {Efstratiou}}, \bibinfo {author} {\bibfnamefont {J.}~\bibnamefont
  {Blackburn}}, \bibinfo {author} {\bibfnamefont {T.}~\bibnamefont
  {Caulfield}}, \bibinfo {author} {\bibfnamefont {G.}~\bibnamefont
  {Stringhini}}, \bibinfo {author} {\bibfnamefont {S.}~\bibnamefont
  {Zannettou}},\ and\ \bibinfo {author} {\bibfnamefont {E.}~\bibnamefont
  {De~Cristofaro}},\ }in\ \href {https://doi.org/10.1609/icwsm.v17i1.22138}
  {\emph {\bibinfo {booktitle} {Proceedings of the Seventeenth International
  AAAI Conference on Web and Social Media (ICWSM2023)}}},\ Vol.~\bibinfo
  {volume} {17}\ (\bibinfo {year} {2023})\ pp.\ \bibinfo {pages}
  {197--208}\BibitemShut {NoStop}%
\bibitem [{\citenamefont {Iyengar}\ \emph {et~al.}(2019)\citenamefont
  {Iyengar}, \citenamefont {Lelkes}, \citenamefont {Levendusky}, \citenamefont
  {Malhotra},\ and\ \citenamefont {Westwood}}]{iyengar2019origins}%
  \BibitemOpen
  \bibfield  {author} {\bibinfo {author} {\bibfnamefont {S.}~\bibnamefont
  {Iyengar}}, \bibinfo {author} {\bibfnamefont {Y.}~\bibnamefont {Lelkes}},
  \bibinfo {author} {\bibfnamefont {M.}~\bibnamefont {Levendusky}}, \bibinfo
  {author} {\bibfnamefont {N.}~\bibnamefont {Malhotra}},\ and\ \bibinfo
  {author} {\bibfnamefont {S.~J.}\ \bibnamefont {Westwood}},\ }\href
  {https://doi.org/10.1146/annurev-polisci-051117-073034} {\bibfield  {journal}
  {\bibinfo  {journal} {Annu. Rev. Polit. Sci.}\ }\textbf {\bibinfo {volume}
  {22}},\ \bibinfo {pages} {129} (\bibinfo {year} {2019})}\BibitemShut
  {NoStop}%
\bibitem [{\citenamefont {Yarchi}\ \emph {et~al.}(2020)\citenamefont {Yarchi},
  \citenamefont {Baden},\ and\ \citenamefont
  {{Kligler-Vilenchik}}}]{yarchi2020political}%
  \BibitemOpen
  \bibfield  {author} {\bibinfo {author} {\bibfnamefont {M.}~\bibnamefont
  {Yarchi}}, \bibinfo {author} {\bibfnamefont {C.}~\bibnamefont {Baden}},\ and\
  \bibinfo {author} {\bibfnamefont {N.}~\bibnamefont {{Kligler-Vilenchik}}},\
  }\href {https://doi.org/10.1080/10584609.2020.1785067} {\bibfield  {journal}
  {\bibinfo  {journal} {Polit. Commun.}\ }\textbf {\bibinfo {volume} {0}},\
  \bibinfo {pages} {1} (\bibinfo {year} {2020})}\BibitemShut {NoStop}%
\bibitem [{\citenamefont {Keuchenius}\ \emph {et~al.}(2021)\citenamefont
  {Keuchenius}, \citenamefont {Törnberg},\ and\ \citenamefont
  {Uitermark}}]{keuchenius2021}%
  \BibitemOpen
  \bibfield  {author} {\bibinfo {author} {\bibfnamefont {A.}~\bibnamefont
  {Keuchenius}}, \bibinfo {author} {\bibfnamefont {P.}~\bibnamefont
  {Törnberg}},\ and\ \bibinfo {author} {\bibfnamefont {J.}~\bibnamefont
  {Uitermark}},\ }\href {https://doi.org/10.1371/journal.pone.0256696}
  {\bibfield  {journal} {\bibinfo  {journal} {PLOS ONE}\ }\textbf {\bibinfo
  {volume} {16}},\ \bibinfo {pages} {1} (\bibinfo {year} {2021})}\BibitemShut
  {NoStop}%
\bibitem [{\citenamefont {Zaslavsky}(1982)}]{zaslavsky1982signed}%
  \BibitemOpen
  \bibfield  {author} {\bibinfo {author} {\bibfnamefont {T.}~\bibnamefont
  {Zaslavsky}},\ }\href
  {https://doi.org/https://doi.org/10.1016/0166-218X(82)90033-6} {\bibfield
  {journal} {\bibinfo  {journal} {Discrete Appl. Math.}\ }\textbf {\bibinfo
  {volume} {4}},\ \bibinfo {pages} {47} (\bibinfo {year} {1982})}\BibitemShut
  {NoStop}%
\bibitem [{\citenamefont {Harary}(1953)}]{harary1953notion}%
  \BibitemOpen
  \bibfield  {author} {\bibinfo {author} {\bibfnamefont {F.}~\bibnamefont
  {Harary}},\ }\href {https://doi.org/10.1307/mmj/1028989917} {\bibfield
  {journal} {\bibinfo  {journal} {Mich. Math. J.}\ }\textbf {\bibinfo {volume}
  {2}},\ \bibinfo {pages} {143} (\bibinfo {year} {1953})}\BibitemShut {NoStop}%
\bibitem [{\citenamefont {Cartwright}\ and\ \citenamefont
  {Harary}(1956)}]{cartwright1956structural}%
  \BibitemOpen
  \bibfield  {author} {\bibinfo {author} {\bibfnamefont {D.}~\bibnamefont
  {Cartwright}}\ and\ \bibinfo {author} {\bibfnamefont {F.}~\bibnamefont
  {Harary}},\ }\href {https://doi.org/10.1037/h0046049} {\bibfield  {journal}
  {\bibinfo  {journal} {Psychol. Rev.}\ }\textbf {\bibinfo {volume} {63}},\
  \bibinfo {pages} {277} (\bibinfo {year} {1956})}\BibitemShut {NoStop}%
\bibitem [{\citenamefont {Neal}(2020)}]{neal2020}%
  \BibitemOpen
  \bibfield  {author} {\bibinfo {author} {\bibfnamefont {Z.~P.}\ \bibnamefont
  {Neal}},\ }\href
  {https://doi.org/https://doi.org/10.1016/j.socnet.2018.07.007} {\bibfield
  {journal} {\bibinfo  {journal} {Soc. Netw.}\ }\textbf {\bibinfo {volume}
  {60}},\ \bibinfo {pages} {103} (\bibinfo {year} {2020})}\BibitemShut
  {NoStop}%
\bibitem [{\citenamefont {Diaz-Diaz}\ and\ \citenamefont
  {Estrada}(2024)}]{diaz2024signed}%
  \BibitemOpen
  \bibfield  {author} {\bibinfo {author} {\bibfnamefont {F.}~\bibnamefont
  {Diaz-Diaz}}\ and\ \bibinfo {author} {\bibfnamefont {E.}~\bibnamefont
  {Estrada}},\ }\href@noop {} {\bibinfo {title} {Signed graphs in data sciences
  via communicability geometry}} (\bibinfo {year} {2024}),\ \Eprint
  {https://arxiv.org/abs/2403.07493} {arXiv:2403.07493} \BibitemShut {NoStop}%
\bibitem [{\citenamefont {Saiz}\ \emph {et~al.}(2017)\citenamefont {Saiz},
  \citenamefont {G{\'o}mez-Garde{\~n}es}, \citenamefont {Nuche}, \citenamefont
  {Gir{\'o}n}, \citenamefont {Pueyo},\ and\ \citenamefont
  {Alados}}]{saiz2017evidence}%
  \BibitemOpen
  \bibfield  {author} {\bibinfo {author} {\bibfnamefont {H.}~\bibnamefont
  {Saiz}}, \bibinfo {author} {\bibfnamefont {J.}~\bibnamefont
  {G{\'o}mez-Garde{\~n}es}}, \bibinfo {author} {\bibfnamefont {P.}~\bibnamefont
  {Nuche}}, \bibinfo {author} {\bibfnamefont {A.}~\bibnamefont {Gir{\'o}n}},
  \bibinfo {author} {\bibfnamefont {Y.}~\bibnamefont {Pueyo}},\ and\ \bibinfo
  {author} {\bibfnamefont {C.~L.}\ \bibnamefont {Alados}},\ }\href
  {https://doi.org/10.1111/ecog.02561} {\bibfield  {journal} {\bibinfo
  {journal} {Ecography}\ }\textbf {\bibinfo {volume} {40}},\ \bibinfo {pages}
  {733} (\bibinfo {year} {2017})}\BibitemShut {NoStop}%
\bibitem [{\citenamefont {Kirkley}\ \emph {et~al.}(2019)\citenamefont
  {Kirkley}, \citenamefont {Cantwell},\ and\ \citenamefont
  {Newman}}]{kirkley2019balance}%
  \BibitemOpen
  \bibfield  {author} {\bibinfo {author} {\bibfnamefont {A.}~\bibnamefont
  {Kirkley}}, \bibinfo {author} {\bibfnamefont {G.~T.}\ \bibnamefont
  {Cantwell}},\ and\ \bibinfo {author} {\bibfnamefont {M.~E.~J.}\ \bibnamefont
  {Newman}},\ }\href {https://doi.org/10.1103/PhysRevE.99.012320} {\bibfield
  {journal} {\bibinfo  {journal} {Phys. Rev. E}\ }\textbf {\bibinfo {volume}
  {99}},\ \bibinfo {pages} {012320} (\bibinfo {year} {2019})}\BibitemShut
  {NoStop}%
\bibitem [{\citenamefont {Facchetti}\ \emph {et~al.}(2011)\citenamefont
  {Facchetti}, \citenamefont {Iacono},\ and\ \citenamefont
  {Altafini}}]{facchetti2011computing}%
  \BibitemOpen
  \bibfield  {author} {\bibinfo {author} {\bibfnamefont {G.}~\bibnamefont
  {Facchetti}}, \bibinfo {author} {\bibfnamefont {G.}~\bibnamefont {Iacono}},\
  and\ \bibinfo {author} {\bibfnamefont {C.}~\bibnamefont {Altafini}},\ }\href
  {https://doi.org/10.1073/pnas.1109521108} {\bibfield  {journal} {\bibinfo
  {journal} {Proc. Natl. Acad. Sci.}\ }\textbf {\bibinfo {volume} {108}},\
  \bibinfo {pages} {20953} (\bibinfo {year} {2011})}\BibitemShut {NoStop}%
\bibitem [{\citenamefont {Gallo}\ \emph {et~al.}(2024)\citenamefont {Gallo},
  \citenamefont {Garlaschelli}, \citenamefont {Lambiotte}, \citenamefont
  {Saracco},\ and\ \citenamefont {Squartini}}]{gallo2024testing}%
  \BibitemOpen
  \bibfield  {author} {\bibinfo {author} {\bibfnamefont {A.}~\bibnamefont
  {Gallo}}, \bibinfo {author} {\bibfnamefont {D.}~\bibnamefont {Garlaschelli}},
  \bibinfo {author} {\bibfnamefont {R.}~\bibnamefont {Lambiotte}}, \bibinfo
  {author} {\bibfnamefont {F.}~\bibnamefont {Saracco}},\ and\ \bibinfo {author}
  {\bibfnamefont {T.}~\bibnamefont {Squartini}},\ }\href
  {https://doi.org/10.1038/s42005-024-01640-7} {\bibfield  {journal} {\bibinfo
  {journal} {Commun. Phys.}\ }\textbf {\bibinfo {volume} {7}},\ \bibinfo
  {pages} {154} (\bibinfo {year} {2024})}\BibitemShut {NoStop}%
\bibitem [{\citenamefont {Harary}(1957)}]{harary1957structural}%
  \BibitemOpen
  \bibfield  {author} {\bibinfo {author} {\bibfnamefont {F.}~\bibnamefont
  {Harary}},\ }\href {https://doi.org/10.1002/bs.3830020403} {\bibfield
  {journal} {\bibinfo  {journal} {Behav. Sci.}\ }\textbf {\bibinfo {volume}
  {2}},\ \bibinfo {pages} {255} (\bibinfo {year} {1957})}\BibitemShut {NoStop}%
\bibitem [{\citenamefont {Wilson}(2012)}]{wilson2012meaningless}%
  \BibitemOpen
  \bibfield  {author} {\bibinfo {author} {\bibfnamefont {P.~H.}\ \bibnamefont
  {Wilson}},\ }in\ \href {https://doi.org/10.1163/9789004226708_003} {\emph
  {\bibinfo {booktitle} {The Projection and Limitations of Imperial Powers,
  1618-1850}}},\ \bibinfo {series} {History of Warfare}, Vol.~\bibinfo {volume}
  {75}\ (\bibinfo  {publisher} {Brill},\ \bibinfo {year} {2012})\ pp.\ \bibinfo
  {pages} {12--33}\BibitemShut {NoStop}%
\bibitem [{\citenamefont {Estrada}\ and\ \citenamefont
  {Benzi}(2014)}]{estrada2014walk}%
  \BibitemOpen
  \bibfield  {author} {\bibinfo {author} {\bibfnamefont {E.}~\bibnamefont
  {Estrada}}\ and\ \bibinfo {author} {\bibfnamefont {M.}~\bibnamefont
  {Benzi}},\ }\href {https://doi.org/10.1103/PhysRevE.90.042802} {\bibfield
  {journal} {\bibinfo  {journal} {Phys. Rev. E}\ }\textbf {\bibinfo {volume}
  {90}},\ \bibinfo {pages} {042802} (\bibinfo {year} {2014})}\BibitemShut
  {NoStop}%
\bibitem [{\citenamefont {Talaga}\ \emph {et~al.}(2023)\citenamefont {Talaga},
  \citenamefont {Stella}, \citenamefont {Swanson},\ and\ \citenamefont
  {Teixeira}}]{talaga2023polarization}%
  \BibitemOpen
  \bibfield  {author} {\bibinfo {author} {\bibfnamefont {S.}~\bibnamefont
  {Talaga}}, \bibinfo {author} {\bibfnamefont {M.}~\bibnamefont {Stella}},
  \bibinfo {author} {\bibfnamefont {T.~J.}\ \bibnamefont {Swanson}},\ and\
  \bibinfo {author} {\bibfnamefont {A.~S.}\ \bibnamefont {Teixeira}},\ }\href
  {https://doi.org/10.1038/s42005-023-01467-8} {\bibfield  {journal} {\bibinfo
  {journal} {Commun. Phys.}\ }\textbf {\bibinfo {volume} {6}},\ \bibinfo
  {pages} {349} (\bibinfo {year} {2023})}\BibitemShut {NoStop}%
\bibitem [{\citenamefont {Diaz-Diaz}\ \emph {et~al.}(2024)\citenamefont
  {Diaz-Diaz}, \citenamefont {Bartesaghi},\ and\ \citenamefont
  {Estrada}}]{diazdiaz2024mathematical}%
  \BibitemOpen
  \bibfield  {author} {\bibinfo {author} {\bibfnamefont {F.}~\bibnamefont
  {Diaz-Diaz}}, \bibinfo {author} {\bibfnamefont {P.}~\bibnamefont
  {Bartesaghi}},\ and\ \bibinfo {author} {\bibfnamefont {E.}~\bibnamefont
  {Estrada}},\ }\href {https://doi.org/10.1007/s12190-024-02204-2} {\bibfield
  {journal} {\bibinfo  {journal} {J. Appl. Math. Comput.}\ ,\ \bibinfo {pages}
  {1}} (\bibinfo {year} {2024})}\BibitemShut {NoStop}%
\bibitem [{\citenamefont {Aref}\ and\ \citenamefont
  {Wilson}(2018)}]{aref2018measuring}%
  \BibitemOpen
  \bibfield  {author} {\bibinfo {author} {\bibfnamefont {S.}~\bibnamefont
  {Aref}}\ and\ \bibinfo {author} {\bibfnamefont {M.~C.}\ \bibnamefont
  {Wilson}},\ }\href {https://doi.org/10.1093/comnet/cnx044} {\bibfield
  {journal} {\bibinfo  {journal} {J. Complex Netw.}\ }\textbf {\bibinfo
  {volume} {6}},\ \bibinfo {pages} {566} (\bibinfo {year} {2018})}\BibitemShut
  {NoStop}%
\bibitem [{\citenamefont {Fraxanet}\ \emph {et~al.}(2024)\citenamefont
  {Fraxanet}, \citenamefont {Pellert}, \citenamefont {Schweighofer},
  \citenamefont {Gómez},\ and\ \citenamefont
  {Garcia}}]{fraxanet2024unpacking}%
  \BibitemOpen
  \bibfield  {author} {\bibinfo {author} {\bibfnamefont {E.}~\bibnamefont
  {Fraxanet}}, \bibinfo {author} {\bibfnamefont {M.}~\bibnamefont {Pellert}},
  \bibinfo {author} {\bibfnamefont {S.}~\bibnamefont {Schweighofer}}, \bibinfo
  {author} {\bibfnamefont {V.}~\bibnamefont {Gómez}},\ and\ \bibinfo {author}
  {\bibfnamefont {D.}~\bibnamefont {Garcia}},\ }\href
  {https://doi.org/10.1093/pnasnexus/pgae276} {\bibfield  {journal} {\bibinfo
  {journal} {PNAS Nexus}\ ,\ \bibinfo {pages} {276}} (\bibinfo {year}
  {2024})}\BibitemShut {NoStop}%
\bibitem [{\citenamefont {Altafini}(2012)}]{altafini2012}%
  \BibitemOpen
  \bibfield  {author} {\bibinfo {author} {\bibfnamefont {C.}~\bibnamefont
  {Altafini}},\ }\href {https://doi.org/10.1371/journal.pone.0038135}
  {\bibfield  {journal} {\bibinfo  {journal} {PLOS ONE}\ }\textbf {\bibinfo
  {volume} {7}},\ \bibinfo {pages} {1} (\bibinfo {year} {2012})}\BibitemShut
  {NoStop}%
\bibitem [{\citenamefont {Shi}\ \emph {et~al.}(2019)\citenamefont {Shi},
  \citenamefont {Altafini},\ and\ \citenamefont {Baras}}]{shi2019dynamics}%
  \BibitemOpen
  \bibfield  {author} {\bibinfo {author} {\bibfnamefont {G.}~\bibnamefont
  {Shi}}, \bibinfo {author} {\bibfnamefont {C.}~\bibnamefont {Altafini}},\ and\
  \bibinfo {author} {\bibfnamefont {J.~S.}\ \bibnamefont {Baras}},\ }\href
  {https://doi.org/10.1137/17M1134172} {\bibfield  {journal} {\bibinfo
  {journal} {SIAM Rev.}\ }\textbf {\bibinfo {volume} {61}},\ \bibinfo {pages}
  {229} (\bibinfo {year} {2019})}\BibitemShut {NoStop}%
\bibitem [{\citenamefont {Fan}\ \emph {et~al.}(2012)\citenamefont {Fan},
  \citenamefont {Wang}, \citenamefont {Li}, \citenamefont {Li},\ and\
  \citenamefont {Jiang}}]{fan2012}%
  \BibitemOpen
  \bibfield  {author} {\bibinfo {author} {\bibfnamefont {P.}~\bibnamefont
  {Fan}}, \bibinfo {author} {\bibfnamefont {H.}~\bibnamefont {Wang}}, \bibinfo
  {author} {\bibfnamefont {P.}~\bibnamefont {Li}}, \bibinfo {author}
  {\bibfnamefont {W.}~\bibnamefont {Li}},\ and\ \bibinfo {author}
  {\bibfnamefont {Z.}~\bibnamefont {Jiang}},\ }\href
  {https://doi.org/10.1088/1742-5468/2012/08/P08003} {\bibfield  {journal}
  {\bibinfo  {journal} {J. Stat. Mech.: Theory Exp.}\ }\textbf {\bibinfo
  {volume} {2012}}\bibinfo  {number} { (08)},\ \bibinfo {pages}
  {P08003}}\BibitemShut {NoStop}%
\bibitem [{\citenamefont {Yin}\ \emph {et~al.}(2021)\citenamefont {Yin},
  \citenamefont {Hu}, \citenamefont {Chen}, \citenamefont {Yuan},\ and\
  \citenamefont {Li}}]{yin2021}%
  \BibitemOpen
\bibfield  {number} {  }\bibfield  {author} {\bibinfo {author} {\bibfnamefont
  {X.}~\bibnamefont {Yin}}, \bibinfo {author} {\bibfnamefont {X.}~\bibnamefont
  {Hu}}, \bibinfo {author} {\bibfnamefont {Y.}~\bibnamefont {Chen}}, \bibinfo
  {author} {\bibfnamefont {X.}~\bibnamefont {Yuan}},\ and\ \bibinfo {author}
  {\bibfnamefont {B.}~\bibnamefont {Li}},\ }\href
  {https://doi.org/10.1109/TKDE.2019.2947421} {\bibfield  {journal} {\bibinfo
  {journal} {IEEE Trans. Knowl. Data Eng.}\ }\textbf {\bibinfo {volume} {33}},\
  \bibinfo {pages} {2208} (\bibinfo {year} {2021})}\BibitemShut {NoStop}%
\bibitem [{\citenamefont {Lee}\ \emph {et~al.}(2023)\citenamefont {Lee},
  \citenamefont {Lee}, \citenamefont {Min},\ and\ \citenamefont
  {Goh}}]{lee2023threshold}%
  \BibitemOpen
  \bibfield  {author} {\bibinfo {author} {\bibfnamefont {K.-M.}\ \bibnamefont
  {Lee}}, \bibinfo {author} {\bibfnamefont {S.}~\bibnamefont {Lee}}, \bibinfo
  {author} {\bibfnamefont {B.}~\bibnamefont {Min}},\ and\ \bibinfo {author}
  {\bibfnamefont {K.-I.}\ \bibnamefont {Goh}},\ }\href
  {https://doi.org/10.1016/j.chaos.2023.113118} {\bibfield  {journal} {\bibinfo
   {journal} {Chaos Soliton Fract.}\ }\textbf {\bibinfo {volume} {168}},\
  \bibinfo {pages} {113118} (\bibinfo {year} {2023})}\BibitemShut {NoStop}%
\bibitem [{\citenamefont {Tian}\ and\ \citenamefont
  {Lambiotte}(2024)}]{tian2024spreading}%
  \BibitemOpen
  \bibfield  {author} {\bibinfo {author} {\bibfnamefont {Y.}~\bibnamefont
  {Tian}}\ and\ \bibinfo {author} {\bibfnamefont {R.}~\bibnamefont
  {Lambiotte}},\ }\href {https://doi.org/10.1137/22M1542325} {\bibfield
  {journal} {\bibinfo  {journal} {SIAM J. Appl. Dyn. Syst.}\ }\textbf {\bibinfo
  {volume} {23}},\ \bibinfo {pages} {50} (\bibinfo {year} {2024})}\BibitemShut
  {NoStop}%
\bibitem [{\citenamefont {Altafini}(2013)}]{altafini2013}%
  \BibitemOpen
  \bibfield  {author} {\bibinfo {author} {\bibfnamefont {C.}~\bibnamefont
  {Altafini}},\ }\href {https://doi.org/10.1109/TAC.2012.2224251} {\bibfield
  {journal} {\bibinfo  {journal} {IEEE Trans. Automat.}\ }\textbf {\bibinfo
  {volume} {58}},\ \bibinfo {pages} {935} (\bibinfo {year} {2013})}\BibitemShut
  {NoStop}%
\bibitem [{\citenamefont {Li}\ \emph {et~al.}(2015)\citenamefont {Li},
  \citenamefont {Chen}, \citenamefont {Wang},\ and\ \citenamefont
  {Zhang}}]{li2015}%
  \BibitemOpen
  \bibfield  {author} {\bibinfo {author} {\bibfnamefont {Y.}~\bibnamefont
  {Li}}, \bibinfo {author} {\bibfnamefont {W.}~\bibnamefont {Chen}}, \bibinfo
  {author} {\bibfnamefont {Y.}~\bibnamefont {Wang}},\ and\ \bibinfo {author}
  {\bibfnamefont {Z.-L.}\ \bibnamefont {Zhang}},\ }\bibfield  {journal}
  {\bibinfo  {journal} {Internet Math.}\ }\textbf {\bibinfo {volume} {11}},\
  \href {https://doi.org/10.1080/15427951.2013.862884}
  {10.1080/15427951.2013.862884} (\bibinfo {year} {2015})\BibitemShut {NoStop}%
\bibitem [{\citenamefont {Pham}\ \emph {et~al.}(2020)\citenamefont {Pham},
  \citenamefont {Kondor}, \citenamefont {Hanel},\ and\ \citenamefont
  {Thurner}}]{pham2020}%
  \BibitemOpen
  \bibfield  {author} {\bibinfo {author} {\bibfnamefont {T.~M.}\ \bibnamefont
  {Pham}}, \bibinfo {author} {\bibfnamefont {I.}~\bibnamefont {Kondor}},
  \bibinfo {author} {\bibfnamefont {R.}~\bibnamefont {Hanel}},\ and\ \bibinfo
  {author} {\bibfnamefont {S.}~\bibnamefont {Thurner}},\ }\href
  {https://doi.org/10.1098/rsif.2020.0752} {\bibfield  {journal} {\bibinfo
  {journal} {J. R. Soc. Interface.}\ }\textbf {\bibinfo {volume} {17}},\
  \bibinfo {pages} {20200752} (\bibinfo {year} {2020})}\BibitemShut {NoStop}%
\bibitem [{\citenamefont {Pham}\ \emph {et~al.}(2022)\citenamefont {Pham},
  \citenamefont {Korbel}, \citenamefont {Hanel},\ and\ \citenamefont
  {Thurner}}]{pham2022}%
  \BibitemOpen
  \bibfield  {author} {\bibinfo {author} {\bibfnamefont {T.~M.}\ \bibnamefont
  {Pham}}, \bibinfo {author} {\bibfnamefont {J.}~\bibnamefont {Korbel}},
  \bibinfo {author} {\bibfnamefont {R.}~\bibnamefont {Hanel}},\ and\ \bibinfo
  {author} {\bibfnamefont {S.}~\bibnamefont {Thurner}},\ }\href
  {https://doi.org/10.1073/pnas.2121103119} {\bibfield  {journal} {\bibinfo
  {journal} {Proc. Natl. Acad. Sci.}\ }\textbf {\bibinfo {volume} {119}},\
  \bibinfo {pages} {e2121103119} (\bibinfo {year} {2022})}\BibitemShut
  {NoStop}%
\bibitem [{\citenamefont {Kempe}\ \emph {et~al.}(2003)\citenamefont {Kempe},
  \citenamefont {Kleinberg},\ and\ \citenamefont {Tardos}}]{kempe2003}%
  \BibitemOpen
  \bibfield  {author} {\bibinfo {author} {\bibfnamefont {D.}~\bibnamefont
  {Kempe}}, \bibinfo {author} {\bibfnamefont {J.}~\bibnamefont {Kleinberg}},\
  and\ \bibinfo {author} {\bibfnamefont {E.}~\bibnamefont {Tardos}},\ }in\
  \href {https://doi.org/10.1145/956750.956769} {\emph {\bibinfo {booktitle}
  {Proceedings of the Ninth ACM SIGKDD International Conference on Knowledge
  Discovery and Data Mining}}},\ \bibinfo {series and number} {KDD '03}\
  (\bibinfo  {publisher} {Association for Computing Machinery},\ \bibinfo
  {address} {New York, NY, USA},\ \bibinfo {year} {2003})\ p.\ \bibinfo {pages}
  {137–146}\BibitemShut {NoStop}%
\bibitem [{\citenamefont {Watts}(2002)}]{watts2002simple}%
  \BibitemOpen
  \bibfield  {author} {\bibinfo {author} {\bibfnamefont {D.~J.}\ \bibnamefont
  {Watts}},\ }\href {https://doi.org/10.1073/pnas.082090499} {\bibfield
  {journal} {\bibinfo  {journal} {Proc. Natl. Acad. Sci.}\ }\textbf {\bibinfo
  {volume} {99}},\ \bibinfo {pages} {5766} (\bibinfo {year}
  {2002})}\BibitemShut {NoStop}%
\bibitem [{\citenamefont {Vaccari}\ \emph {et~al.}(2016)\citenamefont
  {Vaccari}, \citenamefont {Valeriani}, \citenamefont {Barberá}, \citenamefont
  {Jost}, \citenamefont {Nagler},\ and\ \citenamefont {Tucker}}]{vaccari2016}%
  \BibitemOpen
  \bibfield  {author} {\bibinfo {author} {\bibfnamefont {C.}~\bibnamefont
  {Vaccari}}, \bibinfo {author} {\bibfnamefont {A.}~\bibnamefont {Valeriani}},
  \bibinfo {author} {\bibfnamefont {P.}~\bibnamefont {Barberá}}, \bibinfo
  {author} {\bibfnamefont {J.~T.}\ \bibnamefont {Jost}}, \bibinfo {author}
  {\bibfnamefont {J.}~\bibnamefont {Nagler}},\ and\ \bibinfo {author}
  {\bibfnamefont {J.~A.}\ \bibnamefont {Tucker}},\ }\href
  {https://doi.org/10.1177/2056305116664221} {\bibfield  {journal} {\bibinfo
  {journal} {Soc. Media Soc.}\ }\textbf {\bibinfo {volume} {2}},\ \bibinfo
  {pages} {2056305116664221} (\bibinfo {year} {2016})}\BibitemShut {NoStop}%
\bibitem [{\citenamefont {Hills}(2019)}]{hills2019}%
  \BibitemOpen
  \bibfield  {author} {\bibinfo {author} {\bibfnamefont {T.~T.}\ \bibnamefont
  {Hills}},\ }\href {https://doi.org/10.1177/1745691618803647} {\bibfield
  {journal} {\bibinfo  {journal} {Perspect. Psychol. Sci.}\ }\textbf {\bibinfo
  {volume} {14}},\ \bibinfo {pages} {323} (\bibinfo {year} {2019})}\BibitemShut
  {NoStop}%
\bibitem [{\citenamefont {Hosseinmardi}\ \emph {et~al.}(2021)\citenamefont
  {Hosseinmardi}, \citenamefont {Ghasemian}, \citenamefont {Clauset},
  \citenamefont {Mobius}, \citenamefont {Rothschild},\ and\ \citenamefont
  {Watts}}]{hosseinmardi2021}%
  \BibitemOpen
  \bibfield  {author} {\bibinfo {author} {\bibfnamefont {H.}~\bibnamefont
  {Hosseinmardi}}, \bibinfo {author} {\bibfnamefont {A.}~\bibnamefont
  {Ghasemian}}, \bibinfo {author} {\bibfnamefont {A.}~\bibnamefont {Clauset}},
  \bibinfo {author} {\bibfnamefont {M.}~\bibnamefont {Mobius}}, \bibinfo
  {author} {\bibfnamefont {D.~M.}\ \bibnamefont {Rothschild}},\ and\ \bibinfo
  {author} {\bibfnamefont {D.~J.}\ \bibnamefont {Watts}},\ }\href
  {https://doi.org/10.1073/pnas.2101967118} {\bibfield  {journal} {\bibinfo
  {journal} {Proc. Natl. Acad. Sci.}\ }\textbf {\bibinfo {volume} {118}},\
  \bibinfo {pages} {e2101967118} (\bibinfo {year} {2021})}\BibitemShut
  {NoStop}%
\bibitem [{\citenamefont {Adamic}\ and\ \citenamefont
  {Glance}(2005)}]{polblogs}%
  \BibitemOpen
  \bibfield  {author} {\bibinfo {author} {\bibfnamefont {L.~A.}\ \bibnamefont
  {Adamic}}\ and\ \bibinfo {author} {\bibfnamefont {N.}~\bibnamefont
  {Glance}},\ }in\ \href {https://doi.org/10.1145/1134271.1134277} {\emph
  {\bibinfo {booktitle} {Proceedings of the 3rd International Workshop on Link
  Discovery}}},\ \bibinfo {series and number} {LinkKDD '05}\ (\bibinfo
  {publisher} {Association for Computing Machinery},\ \bibinfo {address} {New
  York, NY, USA},\ \bibinfo {year} {2005})\ pp.\ \bibinfo {pages}
  {36--43}\BibitemShut {NoStop}%
\bibitem [{\citenamefont {Ruiz-Garc{\'\i}a}\ \emph {et~al.}(2023)\citenamefont
  {Ruiz-Garc{\'\i}a}, \citenamefont {Ozaita}, \citenamefont {Pereda},
  \citenamefont {Alfonso}, \citenamefont {Bra{\~n}as-Garza}, \citenamefont
  {Cuesta},\ and\ \citenamefont {S{\'a}nchez}}]{ruiz2023triadic}%
  \BibitemOpen
  \bibfield  {author} {\bibinfo {author} {\bibfnamefont {M.}~\bibnamefont
  {Ruiz-Garc{\'\i}a}}, \bibinfo {author} {\bibfnamefont {J.}~\bibnamefont
  {Ozaita}}, \bibinfo {author} {\bibfnamefont {M.}~\bibnamefont {Pereda}},
  \bibinfo {author} {\bibfnamefont {A.}~\bibnamefont {Alfonso}}, \bibinfo
  {author} {\bibfnamefont {P.}~\bibnamefont {Bra{\~n}as-Garza}}, \bibinfo
  {author} {\bibfnamefont {J.~A.}\ \bibnamefont {Cuesta}},\ and\ \bibinfo
  {author} {\bibfnamefont {A.}~\bibnamefont {S{\'a}nchez}},\ }\href
  {https://doi.org/10.1073/pnas.2215041120} {\bibfield  {journal} {\bibinfo
  {journal} {Proc. Natl. Acad. Sci.}\ }\textbf {\bibinfo {volume} {120}},\
  \bibinfo {pages} {e2215041120} (\bibinfo {year} {2023})}\BibitemShut
  {NoStop}%
\bibitem [{\citenamefont {Aref}\ \emph {et~al.}(2020)\citenamefont {Aref},
  \citenamefont {Dinh}, \citenamefont {Rezapour},\ and\ \citenamefont
  {Diesner}}]{aref2020multilevel}%
  \BibitemOpen
  \bibfield  {author} {\bibinfo {author} {\bibfnamefont {S.}~\bibnamefont
  {Aref}}, \bibinfo {author} {\bibfnamefont {L.}~\bibnamefont {Dinh}}, \bibinfo
  {author} {\bibfnamefont {R.}~\bibnamefont {Rezapour}},\ and\ \bibinfo
  {author} {\bibfnamefont {J.}~\bibnamefont {Diesner}},\ }\href
  {https://doi.org/10.1038/s41598-020-71838-6} {\bibfield  {journal} {\bibinfo
  {journal} {Sci. Rep.}\ }\textbf {\bibinfo {volume} {10}},\ \bibinfo {pages}
  {15228} (\bibinfo {year} {2020})}\BibitemShut {NoStop}%
\bibitem [{Note1()}]{Note1}%
  \BibitemOpen
  \bibinfo {note} {Triangles with a positive product of link signs famously
  respect the old adage ``the friend of my friend is my friend, the enemy of my
  friend if my enemy, the enemy of my enemy is my friend''.}\BibitemShut
  {Stop}%
\bibitem [{\citenamefont {Cucuringu}\ \emph {et~al.}(2019)\citenamefont
  {Cucuringu}, \citenamefont {Davies}, \citenamefont {Glielmo},\ and\
  \citenamefont {Tyagi}}]{cucuringu_2019}%
  \BibitemOpen
  \bibfield  {author} {\bibinfo {author} {\bibfnamefont {M.}~\bibnamefont
  {Cucuringu}}, \bibinfo {author} {\bibfnamefont {P.}~\bibnamefont {Davies}},
  \bibinfo {author} {\bibfnamefont {A.}~\bibnamefont {Glielmo}},\ and\ \bibinfo
  {author} {\bibfnamefont {H.}~\bibnamefont {Tyagi}},\ }in\ \href
  {https://proceedings.mlr.press/v89/cucuringu19a.html} {\emph {\bibinfo
  {booktitle} {Proceedings of the Twenty-Second International Conference on
  Artificial Intelligence and Statistics}}},\ \bibinfo {series} {Proceedings of
  Machine Learning Research}, Vol.~\bibinfo {volume} {89},\ \bibinfo {editor}
  {edited by\ \bibinfo {editor} {\bibfnamefont {K.}~\bibnamefont {Chaudhuri}}\
  and\ \bibinfo {editor} {\bibfnamefont {M.}~\bibnamefont {Sugiyama}}}\
  (\bibinfo  {publisher} {PMLR},\ \bibinfo {year} {2019})\ pp.\ \bibinfo
  {pages} {1088--1098}\BibitemShut {NoStop}%
\bibitem [{sup()}]{supplementary}%
  \BibitemOpen
  \href@noop {} {}\bibinfo {note} {See Supplemental Material at [URL will
  inserted by publisher] for modelling and simulation details, mathematical
  proofs, and additional figures, which includes Ref.~[72].}\BibitemShut
  {Stop}%
\bibitem [{\citenamefont {Centola}\ \emph {et~al.}(2007)\citenamefont
  {Centola}, \citenamefont {Eguíluz},\ and\ \citenamefont
  {Macy}}]{Centola2007}%
  \BibitemOpen
  \bibfield  {author} {\bibinfo {author} {\bibfnamefont {D.}~\bibnamefont
  {Centola}}, \bibinfo {author} {\bibfnamefont {V.~M.}\ \bibnamefont
  {Eguíluz}},\ and\ \bibinfo {author} {\bibfnamefont {M.~W.}\ \bibnamefont
  {Macy}},\ }\href {https://doi.org/10.1016/j.physa.2006.06.018} {\bibfield
  {journal} {\bibinfo  {journal} {Physica A Stat. Mech. Appl.}\ }\textbf
  {\bibinfo {volume} {374}},\ \bibinfo {pages} {449} (\bibinfo {year}
  {2007})}\BibitemShut {NoStop}%
\bibitem [{\citenamefont {Centola}(2018)}]{Centola2018}%
  \BibitemOpen
  \bibfield  {author} {\bibinfo {author} {\bibfnamefont {D.}~\bibnamefont
  {Centola}},\ }\href {https://doi.org/10.1126/science.aav1974} {\bibfield
  {journal} {\bibinfo  {journal} {Science}\ }\textbf {\bibinfo {volume}
  {361}},\ \bibinfo {pages} {1320} (\bibinfo {year} {2018})}\BibitemShut
  {NoStop}%
\bibitem [{\citenamefont {Min}\ and\ \citenamefont
  {San~Miguel}(2018)}]{min2018competition}%
  \BibitemOpen
  \bibfield  {author} {\bibinfo {author} {\bibfnamefont {B.}~\bibnamefont
  {Min}}\ and\ \bibinfo {author} {\bibfnamefont {M.}~\bibnamefont
  {San~Miguel}},\ }\href {https://doi.org/10.1038/s41598-018-32643-4}
  {\bibfield  {journal} {\bibinfo  {journal} {Sci. Rep.}\ }\textbf {\bibinfo
  {volume} {8}},\ \bibinfo {pages} {14580} (\bibinfo {year}
  {2018})}\BibitemShut {NoStop}%
\bibitem [{\citenamefont {Centola}(2010)}]{Centola2010}%
  \BibitemOpen
  \bibfield  {author} {\bibinfo {author} {\bibfnamefont {D.}~\bibnamefont
  {Centola}},\ }\href {https://doi.org/10.1126/science.1185231} {\bibfield
  {journal} {\bibinfo  {journal} {Science}\ }\textbf {\bibinfo {volume}
  {329}},\ \bibinfo {pages} {1194} (\bibinfo {year} {2010})}\BibitemShut
  {NoStop}%
\bibitem [{\citenamefont {Karsai}\ \emph {et~al.}(2014)\citenamefont {Karsai},
  \citenamefont {Iniguez}, \citenamefont {Kaski},\ and\ \citenamefont
  {Kert{\'e}sz}}]{karsai2014complex}%
  \BibitemOpen
  \bibfield  {author} {\bibinfo {author} {\bibfnamefont {M.}~\bibnamefont
  {Karsai}}, \bibinfo {author} {\bibfnamefont {G.}~\bibnamefont {Iniguez}},
  \bibinfo {author} {\bibfnamefont {K.}~\bibnamefont {Kaski}},\ and\ \bibinfo
  {author} {\bibfnamefont {J.}~\bibnamefont {Kert{\'e}sz}},\ }\href
  {https://doi.org/10.1098/rsif.2014.0694} {\bibfield  {journal} {\bibinfo
  {journal} {J. R. Soc. Interface.}\ }\textbf {\bibinfo {volume} {11}},\
  \bibinfo {pages} {20140694} (\bibinfo {year} {2014})}\BibitemShut {NoStop}%
\bibitem [{\citenamefont {Cencetti}\ \emph {et~al.}(2023)\citenamefont
  {Cencetti}, \citenamefont {Contreras}, \citenamefont {Mancastroppa},\ and\
  \citenamefont {Barrat}}]{cencetti2023distinguishing}%
  \BibitemOpen
  \bibfield  {author} {\bibinfo {author} {\bibfnamefont {G.}~\bibnamefont
  {Cencetti}}, \bibinfo {author} {\bibfnamefont {D.~A.}\ \bibnamefont
  {Contreras}}, \bibinfo {author} {\bibfnamefont {M.}~\bibnamefont
  {Mancastroppa}},\ and\ \bibinfo {author} {\bibfnamefont {A.}~\bibnamefont
  {Barrat}},\ }\href {https://doi.org/10.1103/PhysRevLett.130.247401}
  {\bibfield  {journal} {\bibinfo  {journal} {Phys. Rev. Lett.}\ }\textbf
  {\bibinfo {volume} {130}},\ \bibinfo {pages} {247401} (\bibinfo {year}
  {2023})}\BibitemShut {NoStop}%
\bibitem [{\citenamefont {{Diaz-Diaz}}\ \emph {et~al.}(2022)\citenamefont
  {{Diaz-Diaz}}, \citenamefont {San~Miguel},\ and\ \citenamefont
  {Meloni}}]{diazdiaz2022}%
  \BibitemOpen
  \bibfield  {author} {\bibinfo {author} {\bibfnamefont {F.}~\bibnamefont
  {{Diaz-Diaz}}}, \bibinfo {author} {\bibfnamefont {M.}~\bibnamefont
  {San~Miguel}},\ and\ \bibinfo {author} {\bibfnamefont {S.}~\bibnamefont
  {Meloni}},\ }\href {https://doi.org/10.1038/s41598-022-13343-6} {\bibfield
  {journal} {\bibinfo  {journal} {Sci. Rep.}\ }\textbf {\bibinfo {volume}
  {12}},\ \bibinfo {pages} {9350} (\bibinfo {year} {2022})}\BibitemShut
  {NoStop}%
\bibitem [{\citenamefont {Abella}\ \emph {et~al.}(2023)\citenamefont {Abella},
  \citenamefont {San~Miguel},\ and\ \citenamefont {Ramasco}}]{abella2023aging}%
  \BibitemOpen
  \bibfield  {author} {\bibinfo {author} {\bibfnamefont {D.}~\bibnamefont
  {Abella}}, \bibinfo {author} {\bibfnamefont {M.}~\bibnamefont {San~Miguel}},\
  and\ \bibinfo {author} {\bibfnamefont {J.~J.}\ \bibnamefont {Ramasco}},\
  }\href {https://doi.org/10.1103/PhysRevE.107.024101} {\bibfield  {journal}
  {\bibinfo  {journal} {Phys. Rev. E}\ }\textbf {\bibinfo {volume} {107}},\
  \bibinfo {pages} {024101} (\bibinfo {year} {2023})}\BibitemShut {NoStop}%
\bibitem [{\citenamefont {P{\'e}rez-Mart{\'\i}nez}\ \emph
  {et~al.}(2023)\citenamefont {P{\'e}rez-Mart{\'\i}nez}, \citenamefont
  {Mingueza}, \citenamefont {Soriano-Pa{\~n}os}, \citenamefont
  {G{\'o}mez-Garde{\~n}es},\ and\ \citenamefont
  {Flor{\'\i}a}}]{perez2023polarized}%
  \BibitemOpen
  \bibfield  {author} {\bibinfo {author} {\bibfnamefont {H.}~\bibnamefont
  {P{\'e}rez-Mart{\'\i}nez}}, \bibinfo {author} {\bibfnamefont {F.~B.}\
  \bibnamefont {Mingueza}}, \bibinfo {author} {\bibfnamefont {D.}~\bibnamefont
  {Soriano-Pa{\~n}os}}, \bibinfo {author} {\bibfnamefont {J.}~\bibnamefont
  {G{\'o}mez-Garde{\~n}es}},\ and\ \bibinfo {author} {\bibfnamefont {L.~M.}\
  \bibnamefont {Flor{\'\i}a}},\ }\href
  {https://doi.org/10.1016/j.chaos.2023.113917} {\bibfield  {journal} {\bibinfo
   {journal} {Chaos Soliton Fract.}\ }\textbf {\bibinfo {volume} {175}},\
  \bibinfo {pages} {113917} (\bibinfo {year} {2023})}\BibitemShut {NoStop}%
\bibitem [{\citenamefont {Kivelä}\ \emph {et~al.}(2014)\citenamefont
  {Kivelä}, \citenamefont {Arenas}, \citenamefont {Barthelemy}, \citenamefont
  {Gleeson}, \citenamefont {Moreno},\ and\ \citenamefont
  {Porter}}]{kivela2014multilayer}%
  \BibitemOpen
  \bibfield  {author} {\bibinfo {author} {\bibfnamefont {M.}~\bibnamefont
  {Kivelä}}, \bibinfo {author} {\bibfnamefont {A.}~\bibnamefont {Arenas}},
  \bibinfo {author} {\bibfnamefont {M.}~\bibnamefont {Barthelemy}}, \bibinfo
  {author} {\bibfnamefont {J.~P.}\ \bibnamefont {Gleeson}}, \bibinfo {author}
  {\bibfnamefont {Y.}~\bibnamefont {Moreno}},\ and\ \bibinfo {author}
  {\bibfnamefont {M.~A.}\ \bibnamefont {Porter}},\ }\href
  {https://doi.org/10.1093/comnet/cnu016} {\bibfield  {journal} {\bibinfo
  {journal} {Journal of Complex Networks}\ }\textbf {\bibinfo {volume} {2}},\
  \bibinfo {pages} {203} (\bibinfo {year} {2014})}\BibitemShut {NoStop}%
\bibitem [{\citenamefont {Vendeville}\ \emph {et~al.}(2022)\citenamefont
  {Vendeville}, \citenamefont {Giovanidis}, \citenamefont {Papanastasiou},\
  and\ \citenamefont {Guedj}}]{vendeville2022opening}%
  \BibitemOpen
  \bibfield  {author} {\bibinfo {author} {\bibfnamefont {A.}~\bibnamefont
  {Vendeville}}, \bibinfo {author} {\bibfnamefont {A.}~\bibnamefont
  {Giovanidis}}, \bibinfo {author} {\bibfnamefont {E.}~\bibnamefont
  {Papanastasiou}},\ and\ \bibinfo {author} {\bibfnamefont {B.}~\bibnamefont
  {Guedj}},\ }in\ \href {https://doi.org/10.1007/978-3-031-21127-0_7} {\emph
  {\bibinfo {booktitle} {International Conference on Complex Networks and Their
  Applications}}}\ (\bibinfo {organization} {Springer},\ \bibinfo {year}
  {2022})\ pp.\ \bibinfo {pages} {74--85}\BibitemShut {NoStop}%
\bibitem [{\citenamefont {Gleeson}(2013)}]{gleeson2013binary}%
  \BibitemOpen
  \bibfield  {author} {\bibinfo {author} {\bibfnamefont {J.~P.}\ \bibnamefont
  {Gleeson}},\ }\href {https://doi.org/10.1103/PhysRevX.3.021004} {\bibfield
  {journal} {\bibinfo  {journal} {Phys. Rev. X}\ }\textbf {\bibinfo {volume}
  {3}},\ \bibinfo {pages} {021004} (\bibinfo {year} {2013})}\BibitemShut
  {NoStop}%
\bibitem [{\citenamefont {Krawiecki}\ and\ \citenamefont
  {Gradowski}(2024)}]{krawiecki2024q}%
  \BibitemOpen
  \bibfield  {author} {\bibinfo {author} {\bibfnamefont {A.}~\bibnamefont
  {Krawiecki}}\ and\ \bibinfo {author} {\bibfnamefont {T.}~\bibnamefont
  {Gradowski}},\ }\href {https://doi.org/10.1103/PhysRevE.109.014302}
  {\bibfield  {journal} {\bibinfo  {journal} {Phys. Rev. E}\ }\textbf {\bibinfo
  {volume} {109}},\ \bibinfo {pages} {014302} (\bibinfo {year}
  {2024})}\BibitemShut {NoStop}%
\bibitem [{\citenamefont {Unicomb}\ \emph {et~al.}(2018)\citenamefont
  {Unicomb}, \citenamefont {I{\~n}iguez},\ and\ \citenamefont
  {Karsai}}]{unicomb2018threshold}%
  \BibitemOpen
  \bibfield  {author} {\bibinfo {author} {\bibfnamefont {S.}~\bibnamefont
  {Unicomb}}, \bibinfo {author} {\bibfnamefont {G.}~\bibnamefont
  {I{\~n}iguez}},\ and\ \bibinfo {author} {\bibfnamefont {M.}~\bibnamefont
  {Karsai}},\ }\href {https://doi.org/10.1038/s41598-018-21261-9} {\bibfield
  {journal} {\bibinfo  {journal} {Sci. Rep.}\ }\textbf {\bibinfo {volume}
  {8}},\ \bibinfo {pages} {3094} (\bibinfo {year} {2018})}\BibitemShut
  {NoStop}%
\bibitem [{\citenamefont {Vendeville}\ \emph {et~al.}(2024)\citenamefont
  {Vendeville}, \citenamefont {Zhou},\ and\ \citenamefont
  {Guedj}}]{vendeville2024discord}%
  \BibitemOpen
  \bibfield  {author} {\bibinfo {author} {\bibfnamefont {A.}~\bibnamefont
  {Vendeville}}, \bibinfo {author} {\bibfnamefont {S.}~\bibnamefont {Zhou}},\
  and\ \bibinfo {author} {\bibfnamefont {B.}~\bibnamefont {Guedj}},\ }\href
  {https://doi.org/10.1103/PhysRevE.109.024312} {\bibfield  {journal} {\bibinfo
   {journal} {Phys. Rev. E}\ }\textbf {\bibinfo {volume} {109}},\ \bibinfo
  {pages} {024312} (\bibinfo {year} {2024})}\BibitemShut {NoStop}%
\end{thebibliography}%


\providecommand{\noopsort}[1]{}\providecommand{\singleletter}[1]{#1}%
\begin{thebibliography}{5}%
\makeatletter
\providecommand \@ifxundefined [1]{%
 \@ifx{#1\undefined}
}%
\providecommand \@ifnum [1]{%
 \ifnum #1\expandafter \@firstoftwo
 \else \expandafter \@secondoftwo
 \fi
}%
\providecommand \@ifx [1]{%
 \ifx #1\expandafter \@firstoftwo
 \else \expandafter \@secondoftwo
 \fi
}%
\providecommand \natexlab [1]{#1}%
\providecommand \enquote  [1]{``#1''}%
\providecommand \bibnamefont  [1]{#1}%
\providecommand \bibfnamefont [1]{#1}%
\providecommand \citenamefont [1]{#1}%
\providecommand \href@noop [0]{\@secondoftwo}%
\providecommand \href [0]{\begingroup \@sanitize@url \@href}%
\providecommand \@href[1]{\@@startlink{#1}\@@href}%
\providecommand \@@href[1]{\endgroup#1\@@endlink}%
\providecommand \@sanitize@url [0]{\catcode `\\12\catcode `\$12\catcode
  `\&12\catcode `\#12\catcode `\^12\catcode `\_12\catcode `\%12\relax}%
\providecommand \@@startlink[1]{}%
\providecommand \@@endlink[0]{}%
\providecommand \url  [0]{\begingroup\@sanitize@url \@url }%
\providecommand \@url [1]{\endgroup\@href {#1}{\urlprefix }}%
\providecommand \urlprefix  [0]{URL }%
\providecommand \Eprint [0]{\href }%
\providecommand \doibase [0]{https://doi.org/}%
\providecommand \selectlanguage [0]{\@gobble}%
\providecommand \bibinfo  [0]{\@secondoftwo}%
\providecommand \bibfield  [0]{\@secondoftwo}%
\providecommand \translation [1]{[#1]}%
\providecommand \BibitemOpen [0]{}%
\providecommand \bibitemStop [0]{}%
\providecommand \bibitemNoStop [0]{.\EOS\space}%
\providecommand \EOS [0]{\spacefactor3000\relax}%
\providecommand \BibitemShut  [1]{\csname bibitem#1\endcsname}%
\let\auto@bib@innerbib\@empty
\bibitem [{\citenamefont {Vendeville}\ \emph {et~al.}(2022)\citenamefont
  {Vendeville}, \citenamefont {Giovanidis}, \citenamefont {Papanastasiou},\
  and\ \citenamefont {Guedj}}]{vendeville2022opening}%
  \BibitemOpen
  \bibfield  {author} {\bibinfo {author} {\bibfnamefont {A.}~\bibnamefont
  {Vendeville}}, \bibinfo {author} {\bibfnamefont {A.}~\bibnamefont
  {Giovanidis}}, \bibinfo {author} {\bibfnamefont {E.}~\bibnamefont
  {Papanastasiou}},\ and\ \bibinfo {author} {\bibfnamefont {B.}~\bibnamefont
  {Guedj}},\ }in\ \href {https://doi.org/10.1007/978-3-031-21127-0_7} {\emph
  {\bibinfo {booktitle} {International Conference on Complex Networks and Their
  Applications}}}\ (\bibinfo {organization} {Springer},\ \bibinfo {year}
  {2022})\ pp.\ \bibinfo {pages} {74--85}\BibitemShut {NoStop}%
\bibitem [{\citenamefont {Vendeville}\ \emph {et~al.}(2024)\citenamefont
  {Vendeville}, \citenamefont {Zhou},\ and\ \citenamefont
  {Guedj}}]{vendeville2024discord}%
  \BibitemOpen
  \bibfield  {author} {\bibinfo {author} {\bibfnamefont {A.}~\bibnamefont
  {Vendeville}}, \bibinfo {author} {\bibfnamefont {S.}~\bibnamefont {Zhou}},\
  and\ \bibinfo {author} {\bibfnamefont {B.}~\bibnamefont {Guedj}},\ }\href
  {https://doi.org/10.1103/PhysRevE.109.024312} {\bibfield  {journal} {\bibinfo
   {journal} {Phys. Rev. E}\ }\textbf {\bibinfo {volume} {109}},\ \bibinfo
  {pages} {024312} (\bibinfo {year} {2024})}\BibitemShut {NoStop}%
\bibitem [{\citenamefont {Kempe}\ \emph {et~al.}(2003)\citenamefont {Kempe},
  \citenamefont {Kleinberg},\ and\ \citenamefont {Tardos}}]{kempe2003}%
  \BibitemOpen
  \bibfield  {author} {\bibinfo {author} {\bibfnamefont {D.}~\bibnamefont
  {Kempe}}, \bibinfo {author} {\bibfnamefont {J.}~\bibnamefont {Kleinberg}},\
  and\ \bibinfo {author} {\bibfnamefont {E.}~\bibnamefont {Tardos}},\ }in\
  \href {https://doi.org/10.1145/956750.956769} {\emph {\bibinfo {booktitle}
  {Proceedings of the Ninth ACM SIGKDD International Conference on Knowledge
  Discovery and Data Mining}}},\ \bibinfo {series and number} {KDD '03}\
  (\bibinfo  {publisher} {Association for Computing Machinery},\ \bibinfo
  {address} {New York, NY, USA},\ \bibinfo {year} {2003})\ p.\ \bibinfo {pages}
  {137–146}\BibitemShut {NoStop}%
\bibitem [{\citenamefont {Gleeson}(2013)}]{gleeson2013binary}%
  \BibitemOpen
  \bibfield  {author} {\bibinfo {author} {\bibfnamefont {J.~P.}\ \bibnamefont
  {Gleeson}},\ }\href {https://doi.org/10.1103/PhysRevX.3.021004} {\bibfield
  {journal} {\bibinfo  {journal} {Phys. Rev. X}\ }\textbf {\bibinfo {volume}
  {3}},\ \bibinfo {pages} {021004} (\bibinfo {year} {2013})}\BibitemShut
  {NoStop}%
\bibitem [{\citenamefont {Adamic}\ and\ \citenamefont
  {Glance}(2005)}]{polblogs}%
  \BibitemOpen
  \bibfield  {author} {\bibinfo {author} {\bibfnamefont {L.~A.}\ \bibnamefont
  {Adamic}}\ and\ \bibinfo {author} {\bibfnamefont {N.}~\bibnamefont
  {Glance}},\ }in\ \href {https://doi.org/10.1145/1134271.1134277} {\emph
  {\bibinfo {booktitle} {Proceedings of the 3rd International Workshop on Link
  Discovery}}},\ \bibinfo {series and number} {LinkKDD '05}\ (\bibinfo
  {publisher} {Association for Computing Machinery},\ \bibinfo {address} {New
  York, NY, USA},\ \bibinfo {year} {2005})\ pp.\ \bibinfo {pages}
  {36--43}\BibitemShut {NoStop}%
\end{thebibliography}%

\end{document}



\title{Echo chamber effects in signed networks: supplementary material}

\author{Antoine Vendeville}
\email{Corresponding author: antoine.vendeville@sciencespo.fr}
\affiliation{m\'edialab, Sciences Po, 75007 Paris, France}
\affiliation{Complex Systems Institute of Paris Île-de-France (ISC-PIF) CNRS, 75013 Paris, France}
\affiliation{Learning Planet Institute, Research Unit Learning Transitions, 75004 Paris, France}
\author{Fernando Diaz-Diaz}
\email{fernandodiaz@ifisc.uib-csic.es}
\affiliation{IFISC (UIB-CSIC), Institute for Cross-Disciplinary Physics and Complex Systems, Campus Universitat de les Illes Balears, 07122 Palma de Mallorca, Spain}

\date{\today}
             
\maketitle




\section{Simulation details}

\subsection{Signed Independent Cascade Model (SICM)}
The simulation process is as follows. To compute $E$ for Erdös-Rényi networks with density $p$ and degree of balance $\tau$, we first generate $m=20$ such networks via the process described in Sec.\ IIA of the main text. Then we perform $2m$ of the SICM with activation probability $\lambda$ on each of these networks, with $m$ simulations seeded in group $a$, and $m$ simulations seeded in group $b$. The number of seeds is chosen to be $10\%$ of the seeding group every time. The remaining agents are initially inactive. For each simulation, we use a breadth-first search algorithm. Contrary to the traditional Independent Cascade, here the specific agent that activates another one matters in the SICM. For example, consider a inactive agent with two neighbors in state $+1$, where one of the neighbors is connected through a positive link and the other through a negative one. The final state of the agent will change depending on which of the agents activates it. Moreover, since the agent cannot update its state once activated, the fist agent that tries to activate has a better chance of propagating its state. To circumvent this issue, we randomize the order in which neighbors try to activate the agent. Finally, we restart the simulation if no agents were activated in a single community. This way, the values of $\rho$ and $E$ are well defined. 


We also compare the results with an asynchronous version of the process. In this version, after choosing the initial seeds, all their neighbors are placed in a queue. At each step, an inactive agent is picked at random from the queue and activates with probability $\lambda$. If it succeeds, it adopts the same state as the agent that triggered this activation. Whenever an agent becomes activated, each of its neighbors are placed in the queue. The process stops when the queue is empty. Picking agents at random from the queue performs double duty. First, consider that multiple seeds may place the same agent in the queue at the start. The state of this agent, if it activates, will depend on the sign of the edge that joins them with the activating seed. By randomizing, we are assured that the activating seed is selected uniformly among all possible ones. Second, this choice does not matter, if we consider that nodes in the queue are endowed with individual exponential clocks of the same rate, and try to activate when the clocks ring. This is a recurrent modeling assumption in queuing models \cite{vendeville2022opening} and in the Voter model \cite{vendeville2024discord}. It allows for a simple treatment of the question of individual updates ordering, which is not straightforward to answer and would otherwise require careful tracking. Asynchronous updates have a clear impact on the results, as the anomalous regimes in the $\tau<0$ region disappear (Figs.~\ref{ssbm_simple_asynchronous}, \ref{compare_asynchronous}).


\subsection{Signed Linear Threshold Model (SLTM)} \label{complex}
We propose the SLTM to adapt the archetypal Threshold Model \cite{kempe2003} for signed networks. As for the SICM, all agents are initially in state $0$, except a few \emph{seeds} in state $+1$, who all belong to the same group. At each step, every inactive agent $i$ is selected. The average influence from its neighbours is computed as: 
\begin{equation}
	\phi_i(t) = \frac{1}{k_i}\sum_{j=1}^{k_i} w_{ij}x_j,
\end{equation}
where $k_i$ is the degree of agent $i$ (independent of the signs of the links), $x_j\in \{-1,0,+1\}$ is the state of agent $j$ and $w_{ij}$ the sign of the link between $i$ and $j$. Notice that both $w_{ij}$ and $x_j$ can be negative, so $\phi_i\in [-1,1].$ We then update the state of $i$ as follows.
\begin{equation}
	x_i \leftarrow 
	\begin{cases}
		+1 & \text{if} \quad \phi_i(t)>T,\\
		+1 & \text{if} \quad \phi_i(t)=T \quad \text{with probability } 1/2,\\
		-1 & \text{if} \quad \phi_i(t)<-T,\\
		-1 & \text{if} \quad \phi_i(t)=-T \quad \text{with probability } 1/2,\\
		0 & \text{otherwise}.
	\end{cases}
\end{equation}
The threshold $T\in(0,1)$ quantifies the difficulty with which information flows through the network. Thus, $1-T$ is akin to $\lambda$ in the SICM, as it measures how easily information travels.  Note that all activations happen simultaneously each step. In Fig.~\ref{compare_complex_asynchronous} we compare the results with an asynchronous version the process, where agents are activated one by one in a random order. This has a clear impact on the results, as the anomalous regimes in the $\tau<0$ region (almost) disappear.

The simulation process is similar as for the SICM. To compute $E$, we first generate $m=20$ such networks via the process described in Sect.~IIA of the main text. Then we perform $2m$ of the SLTM with threshold $T$ on each of these networks, with $m$ simulations seeded in group $a$, and $m$ simulations seeded in group $b$, resulting in a total of $M=800$ realizations. The number of seeds is chosen to be $10\%$ of the seeding group every time.

\subsection{Choice of timesteps}
The quantity of interest $E$ is calculated only from final states of the dynamics, and by essence the time elapsed between two steps of the dynamics has no impact on the result. Thus, there is no difference between discrete or continuous timestep. We chose discrete for the sake of convenience. The choice of continuous timesteps may be useful to study the dynamical aspect of the models.

\section{Mathematical proofs}

\subsection{Triadic balance as a function of noise}  \label{noise_to_balance_proof}
We prove that for any network,
\begin{equation}
    \mathbb{E}[\bal] = (1-2\eta)^3,
\end{equation} 
where $\bal$ is the degree of triadic balance and $\eta$ is the level of noise as defined in Sec.~IIA of the main text. Given a signed network, consider a triangle $T$ in this network. The triangle has its own degree of balance $\bal(T)$. In the absence of noise ($\eta=0$), we have $\bal(T)=1$. If $\eta>0$, each link in this perfectly balanced version of $T$ is flipped with probability $\eta$.
\begin{itemize}
    \item With probability $(1-\eta)^3$ no link is flipped, and $T$ stays balanced.
    \item With probability $3(1-\eta)^2\eta$ exactly one link is flipped, and $T$ becomes antibalanced.
    \item With probability $3(1-\eta)\eta^2$ two links are flipped, and $T$ stays balanced.
    \item With probability: $\eta^3$ all three links are flipped, and $T$ becomes antibalanced. 
\end{itemize}
The expected degree of balance of $T$ after the flips is
\begin{align}
    \mathbb{E}[\bal(T)] &= (1-\eta)^3 -3(1-\eta)^2\eta +3(1-\eta)\eta^2-\eta^3 \\
    &= (1-2\eta)^3. \label{aux1}
\end{align}
The expected degree of balance $\mathbb{E}[\bal]$ of the whole graph is the average of $\mathbb{E}[\bal(T)]$ over all triangles $T$ after the flips, hence $\mathbb{E}[\bal]=(1-2\eta)^3$ as well.

\subsection{The echo chamber effect is a second order moment}
Let $a, b$ denote two complementary subsets of $V$ representing the two groups in the network, and let $f$ be a vector indicating the group of each agent: $f_i=+1$ if $i\in a$, $f_i = -1$ if $i\in b$. Similarly, let $I^\pm$ represent the set of agents with positive and negative state at the end of the contagion process, and let $x$ be a vector defined as follows: $x_i=\pm 1$ if $i\in I^\pm$, $x_i=0$ otherwise. Furthermore, let $s$ denote the set containing the seeds. Based on these definitions, we can express the fraction of 
agents in a final state $\pm 1$, conditioned on the seeds belonging to group $F_s$ and the analyzed agents being active and belonging to group $F_t$, as:
\begin{align}
    \rho_{F_s F_t}^{\pm} = \frac{\vert\left\{i\in F_t:x_i=\pm1, s \subset F_s\right\}\vert}{N_I(F_s,F_t)}
\end{align}
where  $N_I(F_s,F_t)$ denotes the final number of active agents belonging to group $F_t$. Note that this quantity also depends on the seed group $F_s$, as the seed group can affect the efficiency of the contagion process and thus the final number of active agents. 

By the law of large numbers, and assuming independent final states (pair approximation \cite{gleeson2013binary}), the fraction of active agents in a final state $x_i\in\{\pm 1\}$ is approximately equal to the probability of a final state $x_i$ in agent $i$ conditioned on $i$ being active and belonging to a given target group $F_t$, and the set of seeds belonging to group $F_s$:
\begin{align}
    \rho_{F_s F_t}^{\pm} = \frac{\vert\left\{i\in F_t:x_i=\pm1, s \subset F_s\right\}\vert}{N_I(F_s,F_t)} \approx \mathbb{P}[x_i = \pm 1 | x_i \neq 0, s\subset F_s, i\in F_t].
\end{align}

To obtain an unconditioned expectation of the number of active agents on a given state, we employ the law of total expectation, $\mathbb{E}_{X,Y}[X]=\mathbb{E}_Y[\mathbb{E}_X[X|Y]]$, where the subindex indicates the random variable over which the expectation is taken. In this case, the conditional random variable $Y$ corresponds to both the seed and target groups $F_s$ and $F_t$:
\begin{align}
    \mathbb{E}_{x,F_s,F_t}[x_i|x_i \neq 0]& = \mathbb{E}_{F_s}[\mathbb{E}_{F_t}[\mathbb {E}_x[x_i| x_i \neq 0, s\subset F_s, i\in F_t]]] \nonumber \\
    &=\frac{1}{4}\sum_{F_s\in\{a,b\}}\sum_{F_t\in\{a,b\}} \sum_{x_i\in\{\pm 1\}} x_i \mathbb{P}[x_i |x_i \neq 0, s\subset F_s, i\in F_t]
    \nonumber \\ 
     &=\frac{1}{4}\sum_{F_s\in\{a,b\}}\sum_{F_t\in\{a,b\}} \sum_{x_i\in\{\pm 1\}} x_i \rho^{x_i}_{F_sF_t} \nonumber \\ 
     &= \frac{1}{4} \left(\rho^+_{a a} - \rho^-_{a a} + \rho^+_{b a} - \rho^-_{b a} + \rho^+_{a b} - \rho^-_{a b} + \rho^+_{b b} - \rho^-_{b b} \right).
\end{align}
where the factor $1/4$ comes from assuming equally likely group membership, so that
\begin{equation}
    \mathbb{P}(s\subset a)=\mathbb{P}(s\subset b)=\mathbb{P}(i\in a)=\mathbb{P}(i\in b)=1/2.
\end{equation} 
We have thus proved that the expected value of $x_i$ (conditioned on being active) is linearly dependent on the variables $\rho_{ab}^\pm$, accounting for the sign. We can now do a similar argument to relate the echo chamber effect to these variables. To do so, we first define a variable $c_{si}$ indicating whether the seed and target groups agree:
\begin{align}
    c_{si} = \begin{cases}
        +1 \quad \textrm{if} \quad f_s = f_i \\
        -1 \quad \textrm{if} \quad f_s \neq  f_i,
    \end{cases}
\end{align}
where $i$ is an arbitrary agent, $s$ is the set of seeds and all seeds belong to the same group. Alternatively, we can express $c_{si}$ as a product of the groups of $i$ and $s$: $c_{si} = f_s f_i$. Note that, since the group assignations can be random, $c_{si}$ is a random variable and we may define a joint probability of $x_i$ and $c_{si}$.
We now prove that $E$ is a second-order moment of this joint distribution using the methodology employed before: 
\begin{align}
        \mathbb{E}_{x,F_s,F_t}[x_i c_{si}| x_i \neq 0] &= \mathbb{E}_{F_s}[\mathbb{E}_{F_t}[\mathbb{E}_x[x_ic_i | x_i \neq 0, s \subset F_s, i\in F_t]]] \nonumber \\
        &= \frac{1}{4}\sum_{F_s\in\{a,b\}}\sum_{F_t\in\{a,b\}} \sum_{x_i\in\{\pm 1\}}x_i f_i f_s \mathbb{P}[x_i |x_i \neq 0,s \subset F_s, i\in F_t] \nonumber \\ 
        &= \frac{1}{4} \left(\rho^+_{a a} - \rho^-_{a a} - \rho^+_{b a} + \rho^-_{b a} - \rho^+_{a b} + \rho^-_{a b} + \rho^+_{b b} - \rho^-_{b b} \right)
\end{align}
The last expression is the echo chamber effect, so we have proven that $E$ is a second-order moment of the joint distribution of the final states $x_i$ and the seed-target group agreeement variable $c_{si}$.

\subsection{Echo chamber effects for complete networks with maximum infectivity}
For a contagion process in an ER network under the SICM with $p=\lambda=1$, the echo chamber effect can be easily computed as a function of $\eta$. In this case, all nodes become adopters in a single time step, meaning that we simply need to consider the sign of the edge connecting them to the seed. Consider a seed $i$ and an arbitrary node $j$, and assume that $x_i=+1$.  If $i$ and $j$ belong to the same faction, then $x_j=+1$ with probability $1-\eta$ and $x_j=-1$ with probability $\eta$. Thus, $\rho_{aa}^+=\rho_{bb}^+ = 1-\eta$, and $\rho_{aa}^-=\rho_{bb}^- = \eta$. Similarly, if $i$ and $j$ belong to different factions, then $x_j=+1$ with probability $\eta$ and $x_j=-1$ with probability $1-\eta$, so $\rho_{ab}^+=\rho_{ba}^+ = \eta$ and $\rho_{ab}^-=\rho_{ba}^- = 1-\eta$. Substituting in the definition of $E$, one gets $E=1-2\eta $, which according to Eq. \eqref{aux1} implies that $E=\tau^{1/3}.$

\section{Additional figures}

\begin{figure}
    \centering
    \begin{subfigure}[t]{\textwidth}
        \centering
        \includegraphics[width=\textwidth]{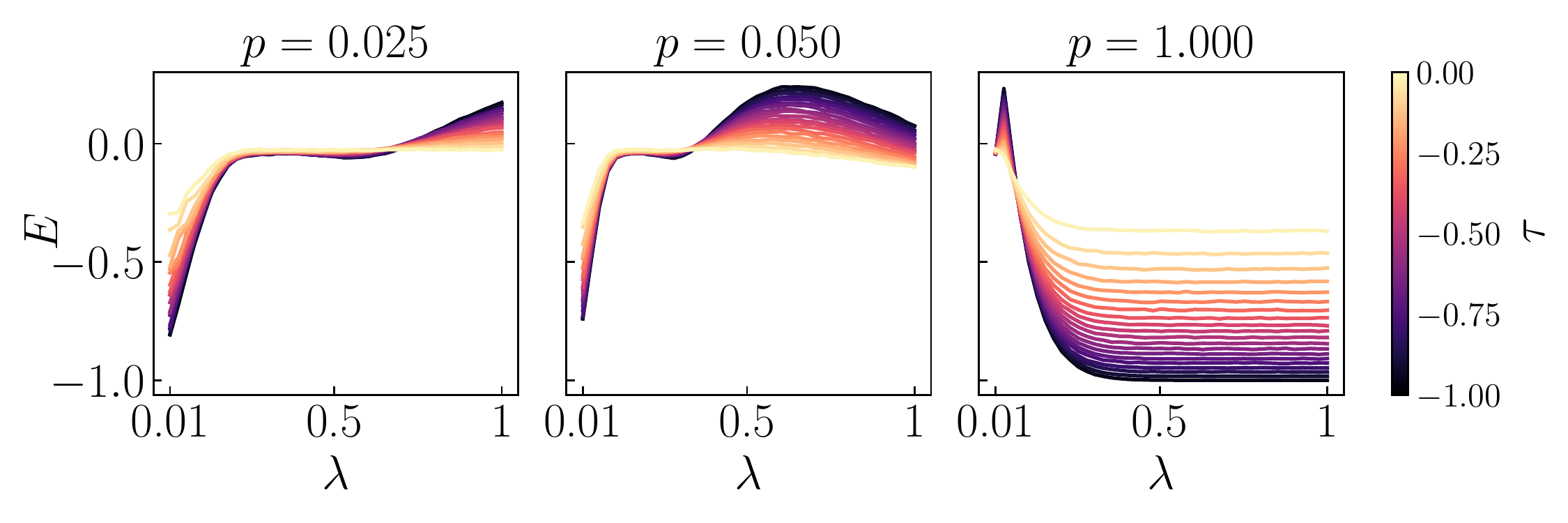}
        \caption{}
        \label{ssbm_simple_neg_lamb}
    \end{subfigure}\\
    \begin{subfigure}[t]{\textwidth}
        \centering
        \includegraphics[width=\textwidth]{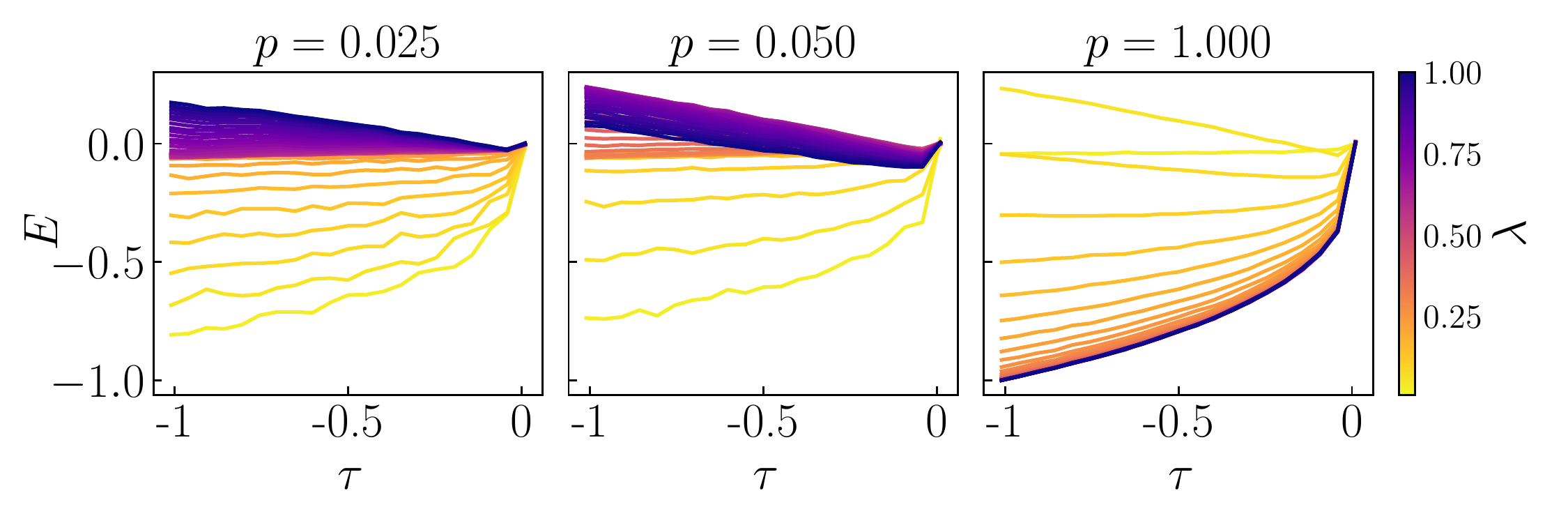}
        \caption{}
        \label{ssbm_simple_neg_tau}
    \end{subfigure}
    \caption{Echo chamber effect in antibalanced ER networks. (a) $E$ as a function of the $\tau$ for various $\lambda$. (b) $E$ as a function of $\lambda$ for various $\tau$. $N=250$ nodes, $M=800$ realisation per data point.}
    \label{ssbm_simple_neg}
\end{figure}

\begin{figure}
	\includegraphics[width=\textwidth]{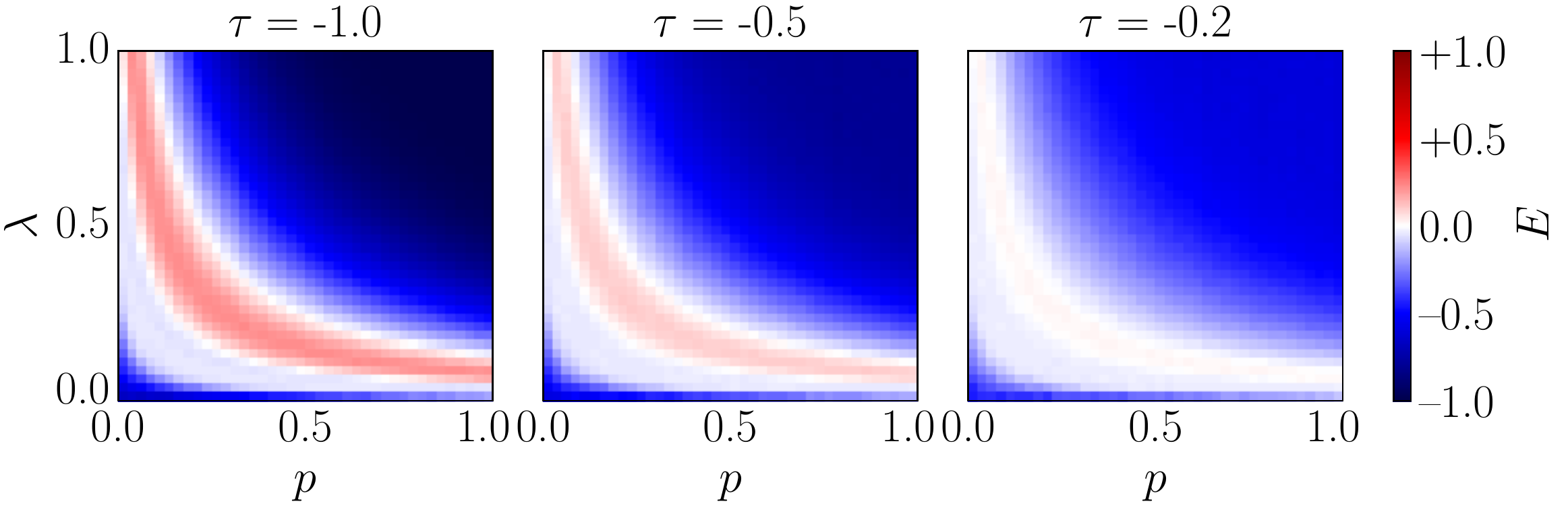}
	\caption{Echo chamber effect as a function of network density and activation probability in antibalanced ER networks with $N=100$ agents, for three different $\tau$. $M=800$ realisation per data point.}
	\label{antibalance100}
\end{figure}

\begin{figure}
    \centering
    \includegraphics[width=\textwidth]{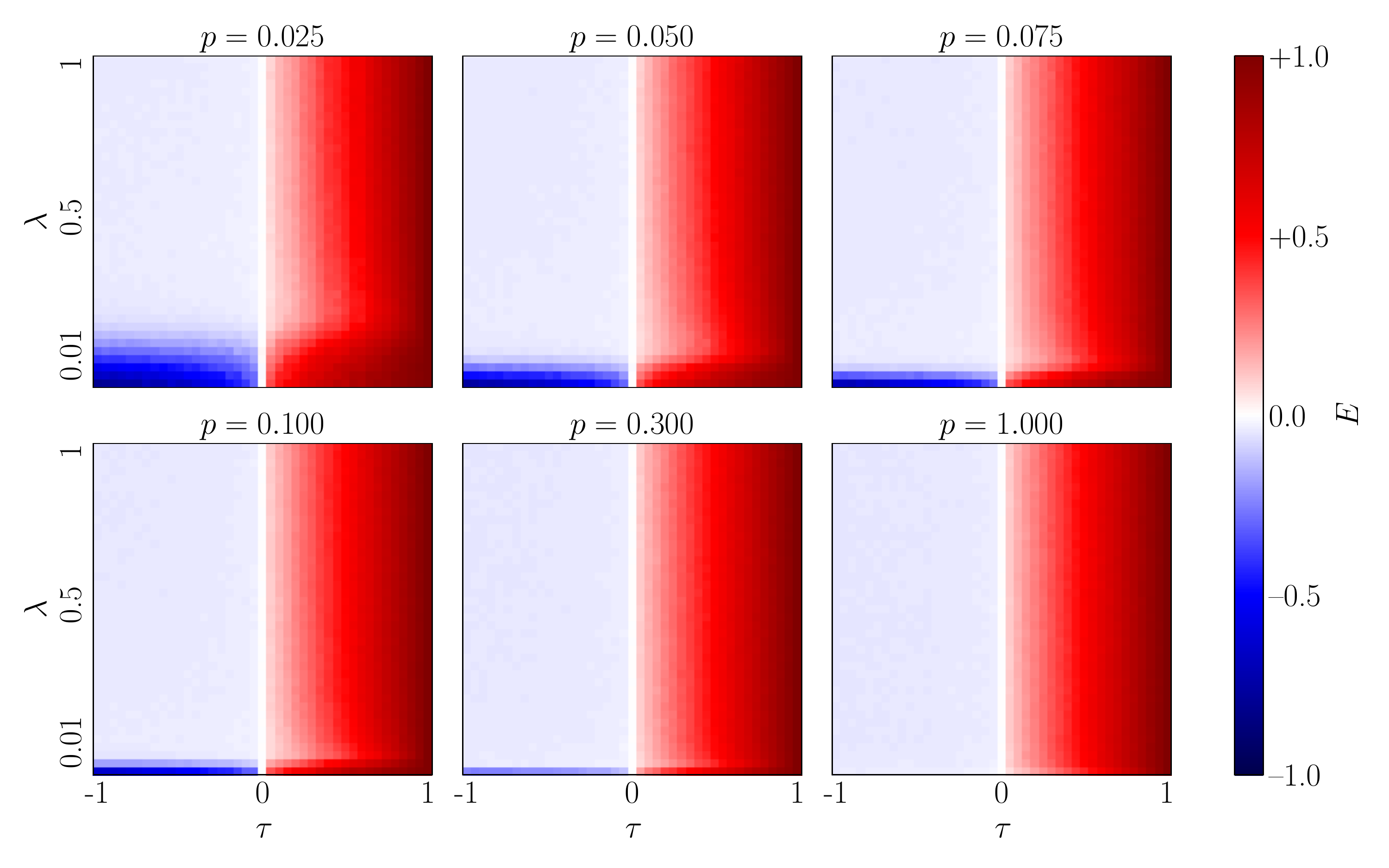}
    \caption{Echo chamber effect in the SICM with asynchronous updates, on ER networks with various densities $p$. $N=250$ nodes, $M=800$ realisation per data point.}
    \label{ssbm_simple_asynchronous}
\end{figure}

\begin{figure}
    \centering
    \includegraphics[width=\textwidth]{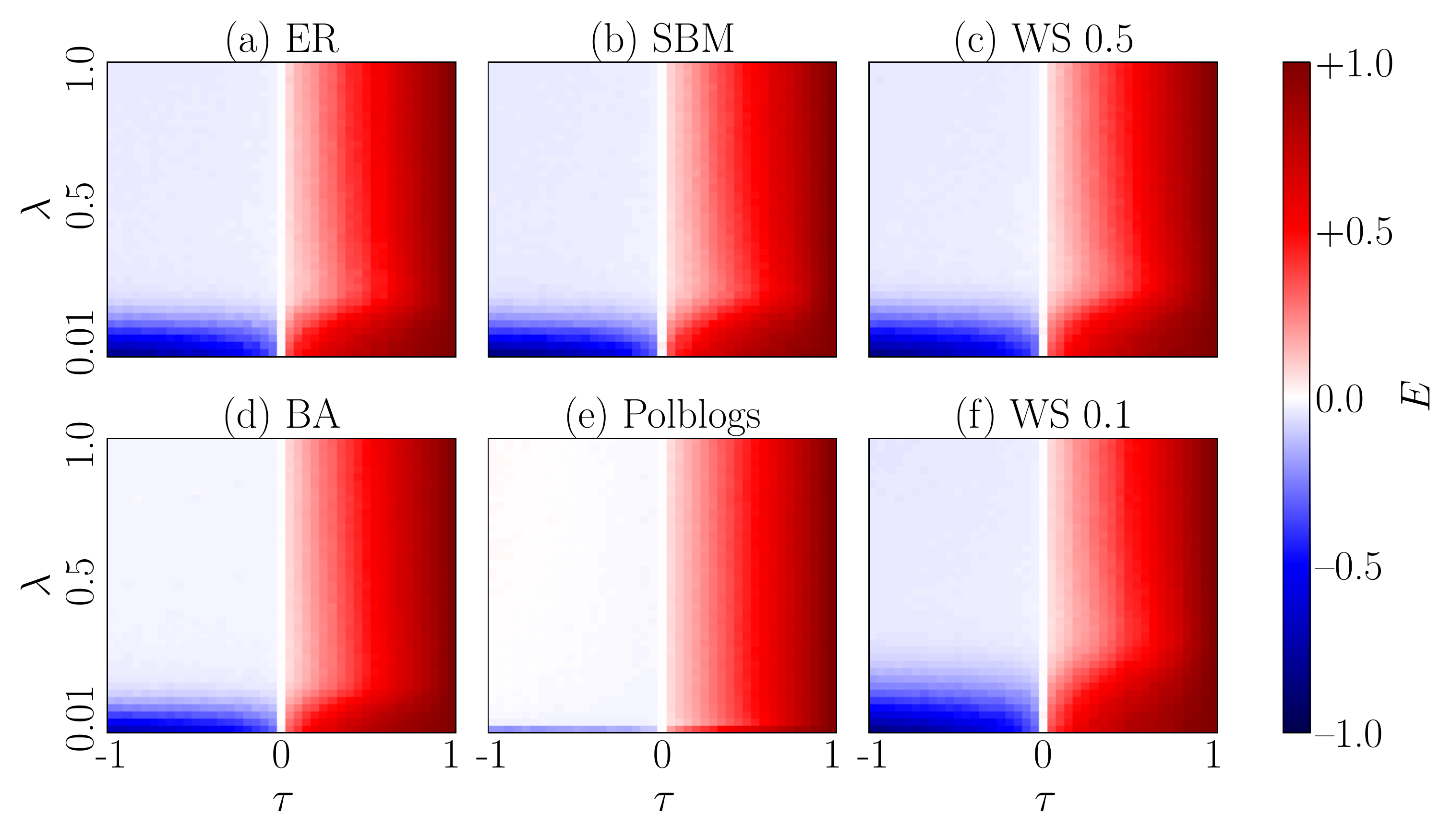}
    \caption{Echo chamber effect in the SICM with asynchronous updates. (a) ER network with $p=0.025$. (b) SBM network with in-group link density $p_{in}=0.4$ and out-group link density $p_{out}=0.1$. Community memberships match group memberships. (c) WS network with average degree $k=6$ and rewiring probability $q=0.5$ (low clustering). (d) BA network with average degree $k=3$. (e) Polblogs network \cite{polblogs}. (f) WS network with average degree $k=6$ and rewiring probability $q=0.1$ (high clustering). All networks have and a density between $0.022$ and $0.025$, and $N=250$ agents---except Polblogs with $N=1,222$.}
    \label{compare_asynchronous}
\end{figure}

\begin{figure}
    \centering
    \includegraphics[width=\textwidth]{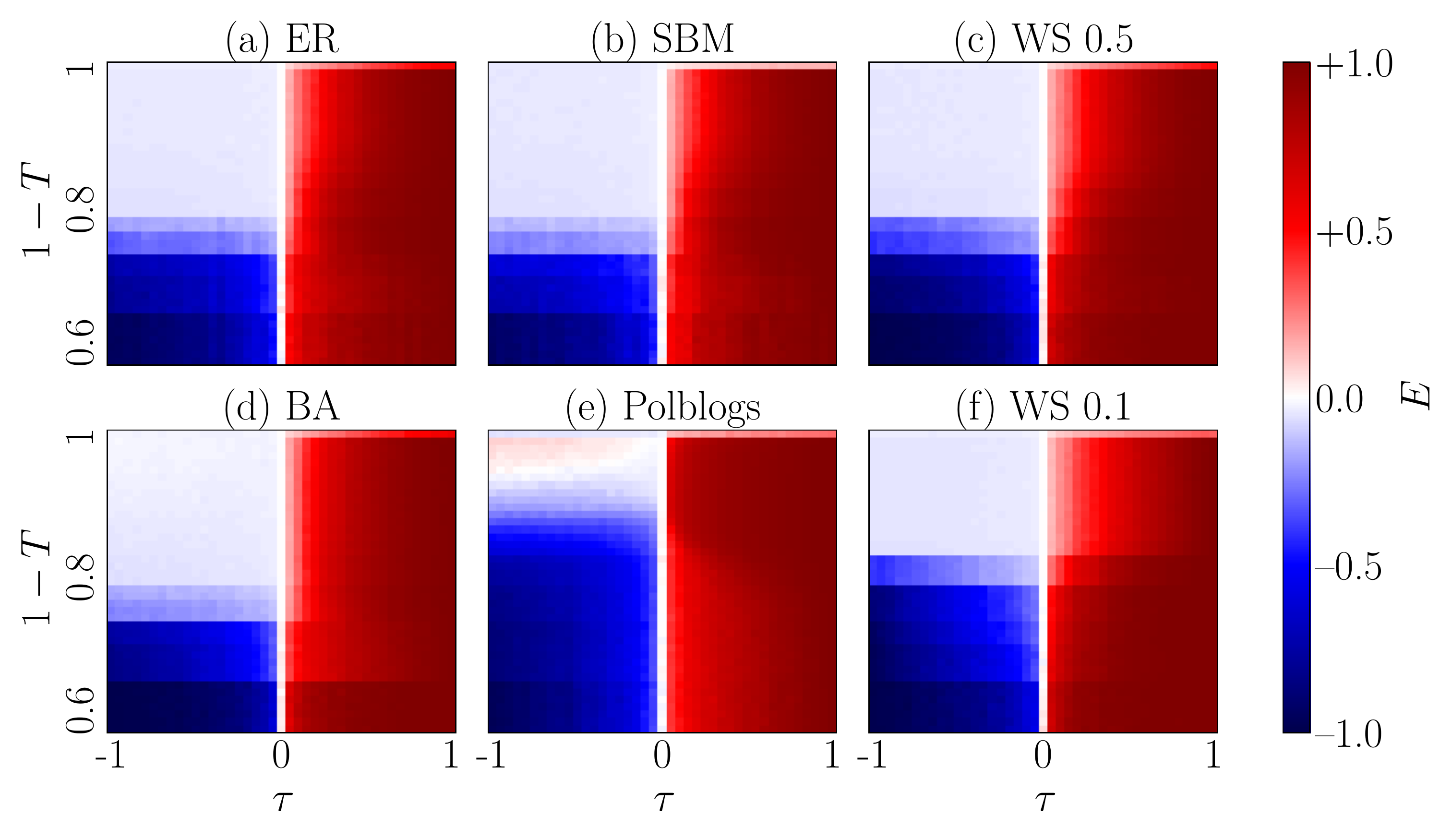}
    \caption{Echo chamber effect in the SLTM with asynchronous updates. (a) ER network with $p=0.025$. (b) SBM network with in-group link density $p_{in}=0.4$ and out-group link density $p_{out}=0.1$. Community memberships match group memberships. (c) WS network with average degree $k=6$ and rewiring probability $q=0.5$ (low clustering). (d) BA network with average degree $k=3$. (e) Polblogs network \cite{polblogs}. (f) WS network with average degree $k=6$ and rewiring probability $q=0.1$ (high clustering). All networks have and a density between $0.022$ and $0.025$, and $N=250$ agents---except Polblogs with $N=1,222$.}
    \label{compare_complex_asynchronous}
\end{figure}

\begin{figure}
    \centering
    \includegraphics[width=\textwidth]{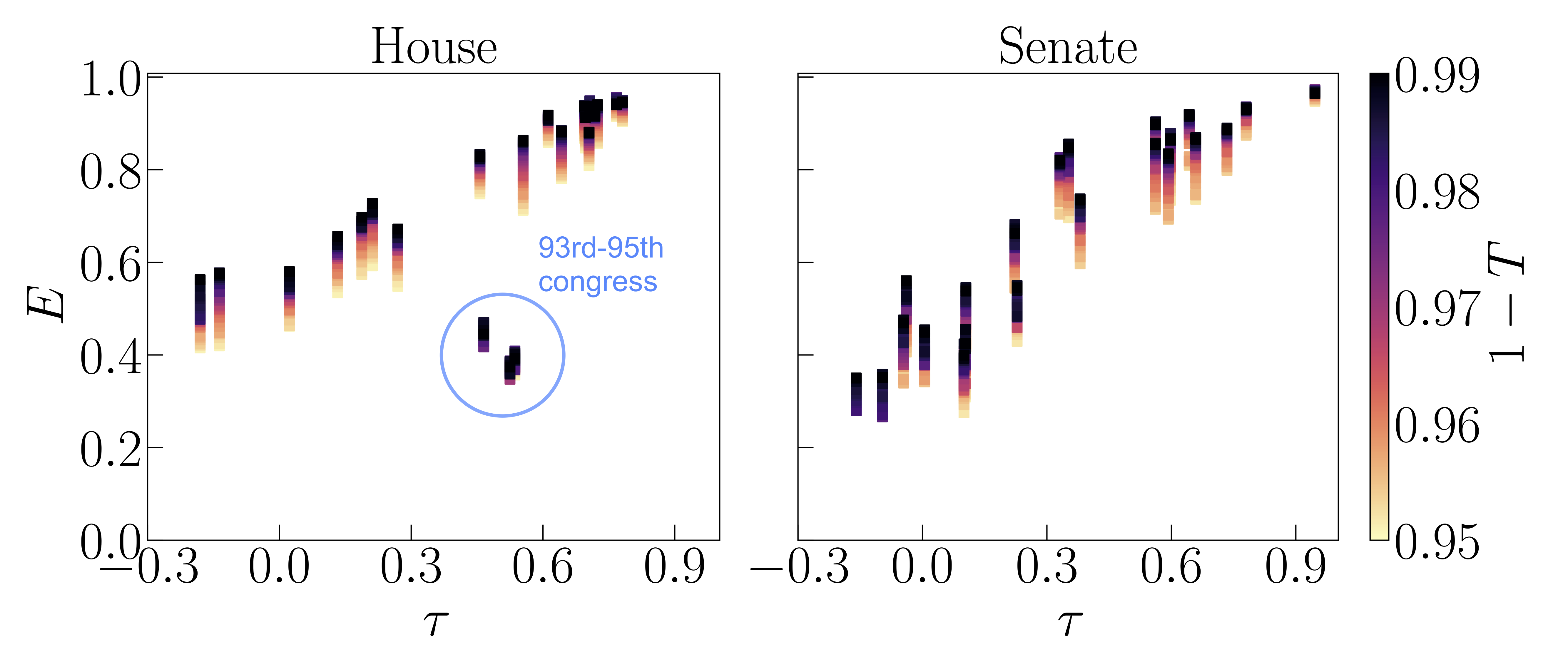}
    \caption{Echo chamber effect $E$ as a function of the degree of balance $\tau$, for SLTM dynamics on the U.S.\ congress networks with various thresholds $T$. We perform $M=100$ simulations per data point.}
    \label{congress_complex}
\end{figure}

Figure~\ref{ssbm_simple_neg} gives another perspective on the highly heterogeneous evolution of $E$ with $\lambda$ and $\tau$ in antibalanced networks and under the SICM. Also for antibalanced networks and SICM, Fig.~\ref{antibalance100} illustrates how the anomalous region changes for simple contagion with a different value of $N$, here $100$ instead of $250$ in the article. The anomalous region has a highly similar shape but is slightly bigger. Figures~\ref{ssbm_simple_asynchronous} and \ref{compare_asynchronous} show that the anomalous region disappear with asynchronous updates in the SICM. In the SLTM they mostly disappear as well, as shown in Fig.~\ref{compare_complex_asynchronous}, except for a tiny one that persists in the Polblogs network. Finally, Fig.~\ref{congress_complex} shows the evolution of the echo chamber effect in the congress network with the SLTM. We obtain mostly the same results as with the SICM, see Fig.\ 7 in the main text.


\bibliographystyle{apsrev4-2}
\bibliography{biblio}